    \documentclass[aps,pre,groupedaddress,amsmath,showpacs,eqsecnum]{revtex4}
   \usepackage{bm}
\usepackage{graphicx,amssymb}
\usepackage{bm}
\usepackage{amsmath}

\newcommand{\beq}{\begin{equation}}
\newcommand{\eeq}{\end{equation}}
\newcommand{\beqa}{\begin{eqnarray}}
\newcommand{\eeqa}{\end{eqnarray}}
\newcommand{\dd}{\text{d}}

\newcommand{\nn}{\nonumber\\}

\newcommand{\el}{\text{el}}

\newcommand{\inel}{\text{inel}}

\newcommand{\TT}{\widetilde{T}}

\newcommand{\nuk}{\chi}

\begin{document}

\title{Granular mixtures modeled as elastic hard spheres subject to a drag force }
\author{Francisco Vega Reyes}
\email{fvega@unex.es}
\author{Vicente Garz\'o}
\email{vicenteg@unex.es}
\homepage{http://www.unex.es/fisteor/vicente/}
\author{Andr\'es Santos}
\email{andres@unex.es} \homepage{http://www.unex.es/fisteor/andres/}
\affiliation{Departamento de F\'{\i}sica, Universidad de
Extremadura, E--06071 Badajoz, Spain}

\date{\today}
\begin{abstract}
Granular  gaseous mixtures  under rapid flow conditions are usually
modeled as a multicomponent system of smooth inelastic hard disks
(two dimensions) or spheres (three dimensions) with constant
coefficients of normal restitution $\alpha_{ij}$. In the low density
regime an adequate framework is provided by the set of coupled
inelastic Boltzmann equations. Due to the intricacy of the inelastic
Boltzmann collision operator, in this paper we propose a simpler
model of elastic hard disks or spheres subject to the action of an
effective drag force, which mimics the effect of dissipation present
in the original granular gas. For each collision term $ij$, the
model has two parameters: a dimensionless factor $\beta_{ij}$
modifying the collision rate of the elastic hard spheres, and the
drag coefficient $\zeta_{ij}$. Both parameters are determined by
requiring that the model reproduces the collisional transfers of
momentum and energy of the true inelastic Boltzmann operator,
yielding $\beta_{ij}=(1+\alpha_{ij})/2$ and $\zeta_{ij}\propto
1-\alpha_{ij}^2$, where the proportionality constant is a function
of the partial densities, velocities, and temperatures of species
$i$ and $j$. The Navier--Stokes transport coefficients for a binary
mixture are obtained from the model by application of the
Chapman--Enskog method. The three coefficients associated with the
mass flux are the same as those obtained from the inelastic
Boltzmann equation, while the remaining four transport coefficients
show a general good agreement, especially in the case of the thermal
conductivity. The discrepancies between both descriptions are seen
to be similar to those found for monocomponent gases. Finally, the
approximate decomposition of the inelastic Boltzmann collision
operator is exploited to construct a model kinetic equation for
granular mixtures as a direct extension of a known kinetic model for
elastic collisions.
\end{abstract}

\maketitle

\section{Introduction\label{sec1}}
Natural and industrial granular media are generally present  in
polydisperse form. In some cases a certain degree of polydispersity
in masses and/or sizes is unavoidable, while in other cases one is
dealing with a real mixture constituted by grains belonging to
 species characterized by distinct mechanical parameters. In
conditions of rapid flow, inelastic binary collisions are the
primary mechanisms and so a kinetic theory description applied to
inelastic hard spheres has proven to be adequate \cite{G03,BP04}. In
the low density regime, all the relevant information on the state of
the mixture is contained in the one-particle velocity distribution
functions $f_i(\mathbf{r},\mathbf{v};t)$, which obey a set of
coupled Boltzmann equations,
\begin{equation}
\left(\partial_t+\mathbf{v}\cdot\nabla\right)f_i=\sum_{j=1}^N J_{ij}^\inel[ {\bf v}|f_i,f_j], \label{n5.1}
\end{equation}
 where $N$ is the number of species and $J_{ij}^\inel[ {\bf v}|f_i,f_j]$ denotes the inelastic Boltzmann
 operator that gives the rate of change of $f_i$ due to collisions
 with particles of species $j$. This collision operator depends parametrically on the coefficient of
 normal restitution $\alpha_{ij}\leq 1$ (here assumed to be constant) for collisions between
particles of species $i$ and $j$.

Obviously, the problem posed by Eq.\ \eqref{n5.1} is much more
complicated than in the case of a single granular gas. Not only one
has to deal with $N$ coupled equations, but in addition the space of
parameters is much larger: there are $N-1$ independent mole
fractions, $N-1$ mass ratios, $N-1$ size ratios, and $N(N-1)/2$
coefficients of restitution. An important consequence of
inelasticity is the breakdown of energy equipartition, even in
homogeneous and isotropic states. This means that one can associate
a different granular temperature to each species
\cite{MP99,GD99b,BT02}, as confirmed by computer simulations
\cite{BT02,MG02,DHGD02,computer,SUKSS06} and real experiments
\cite{SUKSS06,exp1} of agitated mixtures. In the case of small
spatial gradients, the set of Boltzmann equations \eqref{n5.1} can
be solved by means of the Chapman--Enskog method to Navier--Stokes
(NS) order. Many attempts to determine the NS transport coefficients
are restricted to the quasielastic limit ($\alpha_{ij}\approx 1$),
assuming an expansion around Maxwellians at the \emph{same}
temperature \cite{mixture}. A more general derivation takes into
account the
 nonequipartition of energy and determines the transport
 coefficients without any \emph{a priori} limitation on the degree of
 dissipation \cite{GD02}. The accuracy of this latter approach has
 been confirmed by computer simulations in the cases of the diffusion
  \cite{BRCG00,GM04} and  shear viscosity
 \cite{MG03} coefficients.

However, most of the inhomogeneous situations are characterized by a
coupling between inelasticity and spatial gradients, and so they
require a description beyond the NS domain. A typical example of
this coupling is represented by the simple shear flow
\cite{G03,SGD04}, where non-Newtonian effects are unavoidable in the
steady state. Needless to say, the analysis of this type of more
general situations based on the inelastic Boltzmann equation
\eqref{n5.1} becomes very intricate, especially for multicomponent
systems. In order to overcome these difficulties, a possible
strategy consists of replacing the true Boltzmann collision operator
$J_{ij}^\inel$ by a simpler model term that retains its physically
relevant properties. Following an idea previously proposed for
monodisperse granular gases \cite{SA05}, here we explore the
possibility of describing the multicomponent gas of inelastic hard
spheres by a model of elastic hard spheres subject to the action of
an effective drag force with a different drag coefficient for each
species. The parameters of the model are explicitly determined by
optimizing the agreement with the collisional transfer of momentum
and energy obtained from the original operator $J_{ij}^\inel$. The
resulting model is simpler than the original Boltzmann equation
since all the dependence on the coefficients of restitution
$\alpha_{ij}$ appears explicitly outside the collision term.

The paper is organized as follows. In Section \ref{sec2} the model
of elastic hard spheres subject to a drag force is formulated, some
technical details being relegated to  Appendix \ref{appA}. In order
to assess the reliability of the model, in Section \ref{sec3} we
present its Chapman--Enskog solution, the expressions of the NS
transport coefficients being given in Appendix \ref{appB}. In
addition, the dependence of the transport coefficients on
dissipation is compared with  known results derived from the
Boltzmann equation for inelastic hard spheres \cite{GD02,GMD06}. The
paper is closed by a discussion of the results in Section
\ref{sec4}, where a kinetic model equation for granular mixtures is
proposed in Appendix \ref{appC}.

\section{Proposal of the model\label{sec2}}

\subsection{The inelastic Boltzmann equation}

Consider an $N$-component  mixture composed by smooth inelastic
disks ($d=2$) or spheres ($d=3$) of masses $m_{i}$ and diameters
$\sigma _{i}$, $i=1,\ldots,N$. The inelasticity of collisions
between a sphere of species $i$ and a sphere of species $j$  is
characterized by a constant coefficient of restitution $0<\alpha
_{ij}\leq 1$. In the low density regime, the distribution functions
$f_{i}({\bf r},{\bf v};t)$ are determined from the set of nonlinear
Boltzmann equations \eqref{n5.1}, where
 the Boltzmann collision operator  is
\begin{eqnarray}
J_{ij}^\inel\left[ {\bf v}_{1}|f_{i},f_{j}\right] &=&\sigma
_{ij}^{d-1}\int \dd{\bf v} _{2}\int \dd\widehat{\boldsymbol {\sigma
}}\,\Theta (\widehat{{\boldsymbol {\sigma }}} \cdot {\bf
g}_{12})(\widehat{\boldsymbol {\sigma }}\cdot {\bf g}_{12})
\nonumber
\\
&&\times \left[ \alpha _{ij}^{-2}f_{i}({\bf r},{\bf v}_{1}^{\prime
},t)f_{j}( {\bf r},{\bf v}_{2}^{\prime },t)-f_{i}({\bf r},{\bf v}
_{1},t)f_{j}({\bf r}, {\bf v}_{2},t)\right] .
\label{2.2}
\end{eqnarray}
In Eq.\ (\ref{2.2}), $d$ is the dimensionality of the system,
$\sigma _{ij}=\left( \sigma _{i}+\sigma _{j}\right) /2$,
$\widehat{\boldsymbol {\sigma}}$ is a unit vector along the line of
centers, $\Theta $ is the Heaviside step function, and ${\bf
g}_{12}={\bf v}_{1}-{\bf v}_{2}$ is the relative velocity. The
primes on the velocities denote the initial values $\{{\bf
v}_{1}^{\prime }, {\bf v}_{2}^{\prime }\}$ that lead to $\{{\bf
v}_{1},{\bf v}_{2}\}$ following a binary (restituting) collision:
\begin{subequations}
\begin{equation}
{\bf v}_{1}^{\prime }={\bf v}_{1}-\mu _{ji}\left( 1+\alpha
_{ij}^{-1}\right) (\widehat{{\boldsymbol {\sigma }}}\cdot {\bf
g}_{12})\widehat{{\boldsymbol {\sigma }}} ,
\end{equation}
\begin{equation}
 {\bf v}_{2}^{\prime }={\bf v}_{2}+\mu _{ij}\left( 1+\alpha
_{ij}^{-1}\right) (\widehat{{\boldsymbol {\sigma }}}\cdot {\bf
g}_{12})\widehat{ \boldsymbol {\sigma}} ,
\end{equation}
\label{2.3}
\end{subequations}
where $\mu _{ij}\equiv m_{i}/\left( m_{i}+m_{j}\right) $, so that
$\mu_{ij}+\mu_{ji}=1$.

The relevant hydrodynamic fields are the number densities $n_{i}$,
the flow velocity $ {\bf u}$, and the temperature $T$. They are
defined in terms of moments of the distributions $f_{i}$ as
\begin{equation}
n_{i}=\int \dd{\bf v}f_{i}({\bf v}),
\label{2.4.1}
\end{equation}
\begin{equation}
 \rho {\bf
u}=\sum_{i=1}^Nm_{i}n_i \mathbf{u}_i=\sum_{i=1}^Nm_{i} \int \dd {\bf
v}\,{\bf v}f_{i}({\bf v}),
\label{2.4}
\end{equation}
\begin{equation}
nT=p=\sum_{i=1}^N n_i T_i=\sum_{i=1}^2\frac{m_{i}}{d}\int \dd{\bf
v}\,V^{2}f_{i}({\bf v}), \label{2.5}
\end{equation}
where ${\bf V}={\bf v}-{\bf u}$ is the peculiar velocity,
$n=\sum_{i=1}^N n_{i}$ is the total number density,
$\rho=\sum_{i=1}^N\rho_i=\sum_{i=1}^N m_{i}n_{i}$ is the total mass
density, and $p$ is the pressure. Furthermore, the second equality
of Eq.\ \eqref{2.4} and the third equality of Eq.\ (\ref{2.5})
define the flow velocity $\mathbf{u}_i$ and the kinetic temperature
$T_i$ for each species, respectively.

A collision $ij$ conserves the particle number of each species and
the total momentum:
\begin{equation}
\int \dd{\bf v}J_{ij}^\inel[{\bf v}|f_{i},f_{j}]=0,  \label{2.6}
\end{equation}
\begin{equation}
m_i\int \dd{\bf v}\,{\bf v}J_{ij}^\inel[{\bf v}|f_{i},f_{j}]+m_j\int
\dd{\bf v}\,{\bf v}J_{ji}^\inel[{\bf v}|f_{j},f_{i}]=\mathbf{0} .
\label{2.7}
\end{equation}
However, unless $\alpha_{ij}=1$, the collision $ij$ does not
conserve the kinetic energy, so that
\begin{equation}
m_i\int \dd{\bf v}\,v^2J_{ij}^\inel[{\bf v}|f_{i},f_{j}]+m_j\int
\dd{\bf v}\,v^2J_{ji}^\inel[{\bf v}|f_{j},f_{i}]=-\Omega_{ij},
\label{2.7bis}
\end{equation}
where $\Omega_{ij}\geq 0$. The total ``cooling rate'' due to
inelastic collisions among all species is given by
\begin{equation}
\zeta=\frac{1}{2dnT}\sum_{i,j=1}^N \Omega_{ij},
  \label{2.8}
\end{equation}
so that the rate of change of the granular temperature $T$ due to
all the collisions is
\beqa
\left.\ \frac{\partial T}{\partial t}\right|_{\text{coll}}&\equiv&
\frac{1}{nd}\sum_{i,j}m_i\int\dd\mathbf{v} V^2
J_{ij}^\inel[\mathbf{v}|f_i,f_j]\nn &=&-\zeta T.
\eeqa

\subsection{Model of  elastic hard spheres with a drag force}

The dependence of the collision operator $J_{ij}^\inel$ on
$\alpha_{ij}$ is rather involved since it appears as  the factor
$\alpha_{ij}^{-2}$ in the gain term and also through the scattering
rules \eqref{2.3}. This represents an additional difficulty with
respect to the elastic operator $J_{ij}^\el$. In order to simplify
the $\alpha_{ij}$-dependence of the inelastic collision operator, we
propose the following model of elastic particles subject to a drag
force \cite{SA05}:
\begin{equation}
J_{ij}^\inel[{\bf v}|f_i,f_j]\to \beta_{ij} J_{ij}^\el[{\bf
v}|f_i,f_j]+\frac{\zeta_{ij}}{2}\frac{\partial }{\partial {\bf
v}}\cdot \left[\left( {\bf v
}-\mathbf{u}_{i}\right)f_i(\mathbf{v})\right]\equiv K_{ij}[{\bf
v}|f_i,f_j],
\label{5.5}
\end{equation}
where $\beta_{ij}$ and $\zeta_{ij}$ are  determined by optimizing
the agreement between the model and the true operator. The
dimensionless factor $\beta_{ij}$ modifies the collision rate of the
elastic spheres to mimic that of the inelastic spheres. The quantity
$\zeta_{ij}\geq 0$ is the coefficient of the drag force
$\mathbf{F}_{ij}=-(m_i\zeta_{ij}/2) \left( {\bf v
}-\mathbf{u}_{i}\right)$ felt by the elastic spheres of species $i$.
This non-conservative force intends to mimic the loss of energy that
the true inelastic spheres of species $i$ suffer when colliding with
spheres of species $j$. For simplicity, the drag force
$\mathbf{F}_{ij}$ has been chosen proportional to the velocity
relative to the mean flow velocity of species $i$. According to the
model \eqref{5.5}, the Boltzmann equation \eqref{n5.1} becomes
\begin{equation}
\partial_t f_i+\mathbf{v}\cdot\nabla f_i+\frac{1}{m_i}\sum_{j=1}^N
\frac{\partial}{\partial \mathbf{v}}\cdot\left(\mathbf{F}_{ij} f_i
\right)=\sum_{j=1}^N \beta_{ij} J_{ij}^\el[ {\bf v}|f_i,f_j].
\label{n5.1.2}
\end{equation}
In this way, the problem of a mixture of \emph{inelastic} hard
spheres is mapped, via a renormalization of the collision rate and
the introduction of a drag force, onto the problem of a mixture of
\emph{elastic} hard spheres.

The model \eqref{5.5} trivially satisfies the mass conservation
equation \eqref{2.6}. In addition, if we assume the symmetry
relation $\beta_{ij}=\beta_{ji}$ (to be confirmed later), the
momentum conservation equation \eqref{2.7} is also verified.
Finally, Eq.\ \eqref{2.7bis} yields
\beq
n_i\zeta_{ij} \widetilde{T}_i+n_j\zeta_{ji}
\widetilde{T}_j=\frac{1}{d}\Omega_{ij},
\label{n1}
\eeq
where we have introduced the quantity
\begin{eqnarray}
\label{10}
\widetilde{T}_i&=&\frac{m_i}{dn_i}\int \dd{\bf v}\, ({\bf v}-{\bf u}_i)^2f_i\nonumber\\
&=& T_i-\frac{m_i}{d}\left({\bf u}_{i}-{\bf u}\right)^2.
\end{eqnarray}
Note that, according to Eqs.\ \eqref{2.8} and \eqref{n1}, the
cooling rate $\zeta$ of the mixture can be expressed in terms of the
drag coefficients $\zeta_{ij}$ as
\beq
\zeta=\frac{1}{T}\sum_{i,j=1}^N x_i\zeta_{ij}\widetilde{T}_i,
\label{n2}
\eeq
where $x_i=n_i/n$ is the mole fraction of species $i$.

Thus far, the parameters $\beta_{ij}$ and $\zeta_{ij}$ of the model
remain unknown, except for the constraint \eqref{n1}. In order to
determine them, we impose that the collisional transfer of momentum
and energy of species $i$ due to collisions with particles of
species $j$ must be the same as those given by the true Boltzmann
equation,
\begin{equation}
\int \dd{\bf v}\,{\bf v}J_{ij}^\inel[{\bf v}|f_{i},f_{j}]=\int
\dd{\bf v}\,{\bf v}K_{ij}[{\bf v}|f_{i},f_{j}] ,
\label{n2.7}
\end{equation}
\begin{equation}
\int \dd{\bf v}\,v^2J_{ij}^\inel[{\bf v}|f_{i},f_{j}]=\int \dd{\bf
v}\,v^2K_{ij}[{\bf v}|f_{i},f_{j}].
\label{n2.7bis}
\end{equation}
This gives
\begin{equation}
\beta_{ij}=\frac{\int \dd{\bf v}{\bf v}J_{ij}^\inel[{\bf
v}|f_{i},f_{j}]}{\int \dd{\bf v}{\bf v}J_{ij}^\el[{\bf
v}|f_{i},f_{j}]} ,
\label{n3}
\end{equation}
\beq
\zeta_{ij}=\frac{m_i}{dn_i\widetilde{T}_i}\left\{\beta_{ij}\int
\dd{\bf v}v^2J_{ij}^\el[{\bf v}|f_{i},f_{j}]-\int \dd{\bf
v}v^2J_{ij}^\inel[{\bf v}|f_{i},f_{j}]\right\}.
\label{n4}
\eeq
The collision integrals can be simplified by using the property
\cite{BP04}
\beqa
\int \dd{\bf v}_{1}h({\bf v}_{1})J_{ij}^\inel\left[ {\bf
v}_{1}|f_{i},f_{j}\right] &=&\sigma _{ij}^{d-1}\int \dd{\bf
v}_{1}\,\int \dd{\bf v}_{2}f_{i}( {\bf v}_{1})f_{j}({\bf v}_{2})\nn
&&\times \int \dd\widehat{\boldsymbol {\sigma }}\,\Theta
(\widehat{\boldsymbol {\sigma}} \cdot {\bf
g}_{12})(\widehat{\boldsymbol {\sigma }}\cdot {\bf g}_{12})\,\left[
h( {\bf v}_{1}'')-h({\bf v}_{1})\right] ,
\label{c1}
\eeqa
where $h(\mathbf{v})$ is an arbitrary test function and
\begin{equation}
{\bf v}_{1}''={\bf v}_{1}-\mu _{ji}(1+\alpha _{ij})(
\widehat{\boldsymbol {\sigma }}\cdot {\bf
g}_{12})\widehat{\boldsymbol {\sigma}}
\label{c2}
\end{equation}
is the post-collisional velocity of a particle of species $i$. Taking $h(\mathbf{v})=\mathbf{v}$ in Eq.\
\eqref{c1}, Eq.\ \eqref{n3} simply becomes
\begin{equation}
\beta_{ij}=\frac{1+\alpha_{ij}}{2}, \label{n5} \end{equation}
 in agreement with the symmetry relation
$\beta_{ij}=\beta_{ji}$. Next, taking $h(\mathbf{v})=v^2$ and making use of
\begin{equation}
{v_1''}^2-v_1^2=-\mu _{ji}(1+\alpha _{ij})( \widehat{\boldsymbol {\sigma }}\cdot {\bf g}_{12})\left[2(
\widehat{\boldsymbol {\sigma }}\cdot {\bf G}_{12})+\mu _{ji}(1-\alpha _{ij})( \widehat{\boldsymbol {\sigma
}}\cdot {\bf g}_{12})\right], \label{n6}
\end{equation}
where $\mathbf{G}_{12}=\mu_{ij}\mathbf{v}_1+\mu_{ji}\mathbf{v}_2$ is the center of mass velocity, one gets from
Eq.\ \eqref{n4}
\begin{equation}
\zeta_{ij}=\frac{\pi^{(d-1)/2}}{d\Gamma((d+3)/2)}(1-\alpha_{ij}^2)\frac{n_j
m_i\mu_{ji}^2\sigma_{ij}^{d-1}}{\widetilde{T}_i }\langle g_{12}^3\rangle_{ij}.
\label{n7}
\end{equation}
Here,
\begin{equation}
\langle g_{12}^3\rangle_{ij}=\frac{1}{n_i n_j}\int \dd\mathbf{v}_1\int \dd\mathbf{v}_2\, f_i(\mathbf{v}_1)
f_j(\mathbf{v}_2) g_{12}^3
\label{n8}
\end{equation}
and use has been made of the result \cite{NE98}
\begin{equation} \int \dd\widehat{\boldsymbol {\sigma }}\,\Theta
(\widehat{\boldsymbol {\sigma}} \cdot {\bf g}_{12})(\widehat{\boldsymbol {\sigma }}\cdot {\bf
g}_{12})^3=\frac{\pi^{(d-1)/2}}{\Gamma((d+3)/2)}g_{12}^3.
\label{n9}
\end{equation}
Inserting Eq.\ \eqref{n7} into Eq.\ \eqref{n1} we get
\begin{equation}
\Omega_{ij}=\frac{\pi^{(d-1)/2}}{\Gamma((d+3)/2)}(1-\alpha_{ij}^2)n_i n_j\frac{m_i m_{j}}{m_i+m_j
}\sigma_{ij}^{d-1}\langle g_{12}^3\rangle_{ij}.
\label{n10}
\end{equation}
According to Eq.\ \eqref{n7},  the drag coefficient $\zeta_{ij}$ is
a positive definite quantity which only vanishes if the $ij$
collisions are elastic ($\alpha_{ij}=1$). It can be interpreted as
the cooling rate of species $i$ due to the inelasticity of
collisions with particles of species $j$.  To make this more
explicit, let us consider the contribution of collisions with
particles of species $j$ to the rate of change of the partial
temperature $\TT_i$, i.e.,
\begin{eqnarray}
 \left.\frac{\partial\TT_i}{\partial t}\right|_{\text{coll},j}&\equiv&
\frac{m_i}{dn_i} \int \dd\mathbf{v}\, (\mathbf{v}-\mathbf{u}_i)^2
J_{ij}^\inel\left[ {\bf v}|f_{i},f_{j}\right]\nn
&=&\beta_{ij}\frac{m_i}{dn_i}\int \dd\mathbf{v}\,
(\mathbf{v}-\mathbf{u}_i)^2 J_{ij}^\el\left[ {\bf
v}|f_{i},f_{j}\right]-\zeta_{ij}\TT_{i}, \label{new}
\end{eqnarray}
where in the last step use has been made of Eqs.\ \eqref{n2.7} and \eqref{n2.7bis}. The first term on the
right-hand side represents the contribution to the rate of change not directly associated with inelasticity
(except for the presence of the factor $\beta_{ij}$). This term can be either positive or negative, depending on
$f_i$ and $f_j$. In particular, if $\mathbf{u}_i=\mathbf{u}_j$ and $f_i$ and $f_j$ are approximated by
Maxwellians, the sign of this term is the same as that of the temperature difference $\TT_j-\TT_i$. Therefore,
the second term on the right-hand side of \eqref{new} is the genuine contribution associated with inelasticity.
It must be emphasized that Eq.\ \eqref{new}, with $\beta_{ij}$ and $\zeta_{ij}$ given by Eqs.\ \eqref{n5} and
\eqref{n7}, respectively, is exact and so it is not restricted to the model \eqref{5.5}.

While Eq.\ \eqref{n7} is formally exact, it involves the average $\langle g_{12}^3\rangle_{ij}$, which is a
functional of $f_i$ and $f_j$. An estimate of this average can be obtained by assuming Gaussian forms for $f_i$
and $f_j$ given by
\begin{equation}
f_i(\mathbf{v})=n_i\left(\frac{m_i}{2\pi\widetilde{T}_i}\right)^{d/2}
\exp\left[-\frac{m_i(\mathbf{v}-\mathbf{u}_i)^2}{2\widetilde{T}_i}\right]
\label{A1}
\end{equation}
with an analogous form for $f_j(\mathbf{v})$. The distribution
\eqref{A1} is the one that, sharing with the exact distribution the
first $d+2$ velocity moments, maximizes the missing information
defined as $-\int \dd\mathbf{v}\, f_i(\mathbf{v})\ln
f_i(\mathbf{v})$. The corresponding average $\langle
g_{12}^3\rangle_{ij}$ is obtained in Appendix \ref{appA} by
neglecting terms of order fourth and higher in the difference
$\mathbf{u}_i-\mathbf{u}_j$. Inserting Eq.\ \eqref{n11} into Eq.\
\eqref{n7}, one obtains
\begin{equation}
\label{4}
\zeta_{ij}=\frac{1}{2}\xi_{ij}\mu_{ji}^2\left[1+\frac{m_i\widetilde{T}_j}{m_j
\widetilde{T}_i}+\frac{3}{2d}\frac{m_i}{\widetilde{T}_i}\left({\bf
u}_i-{\bf u}_j\right)^2\right](1-\alpha_{ij}^2),
\end{equation}
where
\begin{equation}
\label{5}
\xi_{ij}=\frac{4\pi^{(d-1)/2}}{d\Gamma(d/2)}n_j\sigma_{ij}^{d-1}\left(\frac{2\widetilde{T}_i}{m_i}+
\frac{2\widetilde{T}_j}{m_j}\right)^{1/2}
\end{equation}
is an effective collision frequency of species $i$ due to collisions
with particles of species $j$. Note that $n_i\xi_{ij}=n_j\xi_{ji}$,
while $m_i n_i\TT_i \zeta_{ij}=m_j n_j\TT_j \zeta_{ji}$. Equation
\eqref{4} reduces to the one derived in Ref.\ \onlinecite{SA05} when
$\mathbf{u}_i=\mathbf{u}_j$.

In summary, the model is defined by the replacement \eqref{5.5} with the parameters $\beta_{ij}$ and
$\zeta_{ij}$ given by Eqs.\ \eqref{n5} and \eqref{4}, respectively. While $\beta_{ij}$ depends only on the
coefficient of restitution $\alpha_{ij}$, the cooling rate $\zeta_{ij}$ is also a function of the masses and
sizes of particles of species $i$ and $j$, as well as of the first few velocity moments (density, flow velocity,
and partial temperature) of both distributions $f_i$ and $f_j$. The expressions of $\beta_{ij}$ and $\zeta_{ij}$
preserve the collisional transfer of momentum and energy of the true inelastic Boltzmann equation, although in
the latter case we have considered the Gaussian forms (\ref{A1}) for $f_i$ and $f_j$ parametrized by their first
$d+2$ moments in order to get explicit results.

It is worth mentioning that, while the model \eqref{5.5} is
mathematically simpler than the original Boltzmann operator, its
functional dependence on $f_i$ and $f_j$ through the corresponding
flow velocities and partial temperatures is highly nonlinear.
Therefore, the bilinear property
\beq
J_{ij}^\inel[\mathbf{v}|\lambda_i f_i,\lambda_j
f_j]=\lambda_i\lambda_j J_{ij}^\inel[\mathbf{v}| f_i,f_j]
\label{6}
\eeq
is satisfied, but not the other bilinear property
\beq
J_{ij}^\inel[\mathbf{v}|f_{i1}+f_{i2},f_{j1}+f_{j2}]=J_{ij}^\inel[\mathbf{v}|f_{i1},f_{j1}]+
J_{ij}^\inel[\mathbf{v}|f_{i1},f_{j2}]+J_{ij}^\inel[\mathbf{v}|f_{i2},f_{j1}]
+J_{ij}^\inel[\mathbf{v}|f_{i2},f_{j2}].
\label{7}
\eeq
A consequence of the failure of $K_{ij}$ to satisfy the property
\eqref{7} is that, in general, one does not get a closed equation
for the total distribution function $f=\sum_i f_i$ in the case of
mechanically equivalent particles, unless $\mathbf{u}_i=\mathbf{u}$
and $T_i=T$. However, this drawback is not relevant in most of the
situations of physical interest, such as nonequilibrium steady
states, since in those cases the existence of different velocities
and/or temperatures is a consequence of the particles being
mechanically different.

\subsection{Homogeneous cooling state}

The simplest application of the model corresponds to the so-called
homogeneous cooling state (HCS) \cite{GD99b}. This state is
characterized by  the absence of gradients, so that
$\mathbf{u}_i=\mathbf{u}$. In that case, the model \eqref{5.5}
yields
\beq
\partial_t T_i=-\zeta_i T_i,
\label{n12}
\eeq
where
\beq
\zeta_i=\sum_{j=1}^N \left(-\frac{m_i}{dn_i T_i}\beta_{ij}\int
\dd\mathbf{v} V^2 J_{ij}^\el [\mathbf{v}|f_i,f_j]+\zeta_{ij}\right).
\label{n13}
\eeq
As said above in connection with Eq.\ \eqref{new}, Eq.\ \eqref{n13}
is also valid for the inelastic Boltzmann equation, provided that
$\zeta_{ij}$ is given by Eq.\ \eqref{n7}.

To evaluate the collision integral in Eq.\ \eqref{n13}, we can take
again Eq.\ \eqref{A1}, which  (since $\mathbf{u}_i=\mathbf{u}$)
becomes
\begin{equation}
\label{a11}
f_i\to f_{i,M}=n_i \left(\frac{m_i}{2\pi T_i}\right)^{d/2}
\exp(-m_iV^2/2T_i).
\end{equation}
This yields
\beq
-\frac{m_i}{dn_i T_i}\int \dd\mathbf{v} V^2 J_{ij}^\el
[\mathbf{v}|f_i,f_j]=2\xi_{ij}\frac{m_im_j}{(m_i+m_j)^2}\frac{T_i-T_j}{T_i}.
\label{n14}
\eeq
Therefore,
\beq
\zeta_i=\sum_{j=1}^N
\xi_{ij}\frac{m_im_j}{(m_i+m_j)^2}(1+\alpha_{ij})\left[\frac{T_i-T_j}{T_i}+
\frac{1-\alpha_{ij}}{2}\left(\frac{m_j}{m_i}+\frac{T_j}{T_i}\right)\right].
\label{n15}
\eeq
This expression coincides with the one obtained from the original
Boltzmann equation when the approximation \eqref{a11} is used.

 The HCS condition is $\partial_t T_i/T=0$, which
implies $\zeta_1=\zeta_2=\cdots=\zeta_N$. This gives the $N-1$
temperature ratios $\gamma_i\equiv T_i/T$ as functions of the mole
fractions, the mass ratios, the size ratios, and the coefficients of
restitution of the mixture. Comparison with computer simulations
\cite{BT02,MG02,DHGD02} shows an excellent agreement with the
results obtained from Eq.\ \eqref{n15}, even for strong dissipation.

\section{Navier--Stokes transport coefficients of the
model\label{sec3}}

\begin{figure}
\includegraphics[width=.5\columnwidth]{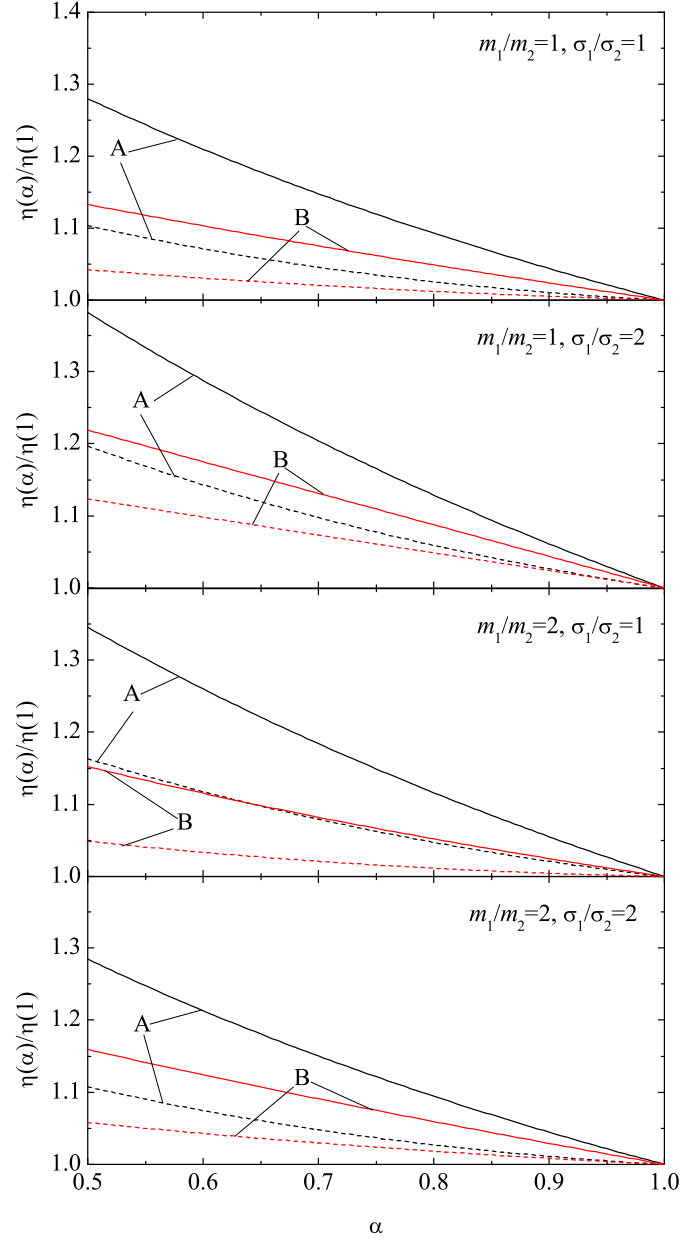}
\caption{Plot of the reduced shear viscosity coefficient
$\eta(\alpha)/\eta(1)$ as a function of the  coefficient of
restitution $\alpha$ for a three-dimensional equimolar mixture in
the cases $\alpha_{11}=\alpha_{12}=\alpha_{22}=\alpha$ (case A) and
$\alpha_{11}=\alpha$, $\alpha_{12}=(1+\alpha)/2$,
$\alpha_{22}=(3+\alpha)/4$ (case B). The panels correspond, from top
to bottom, to the systems $\{m_1/m_2,\sigma_1/\sigma_2\}=\{1,1\}$,
$\{1,2\}$, $\{2,1\}$, and $\{2,2\}$, respectively. The solid lines
are the Boltzmann results, while the dashed lines are the
predictions of the model \protect\eqref{5.5}.}
\label{eta}
\end{figure}

\begin{figure}
\includegraphics[width=.5\columnwidth]{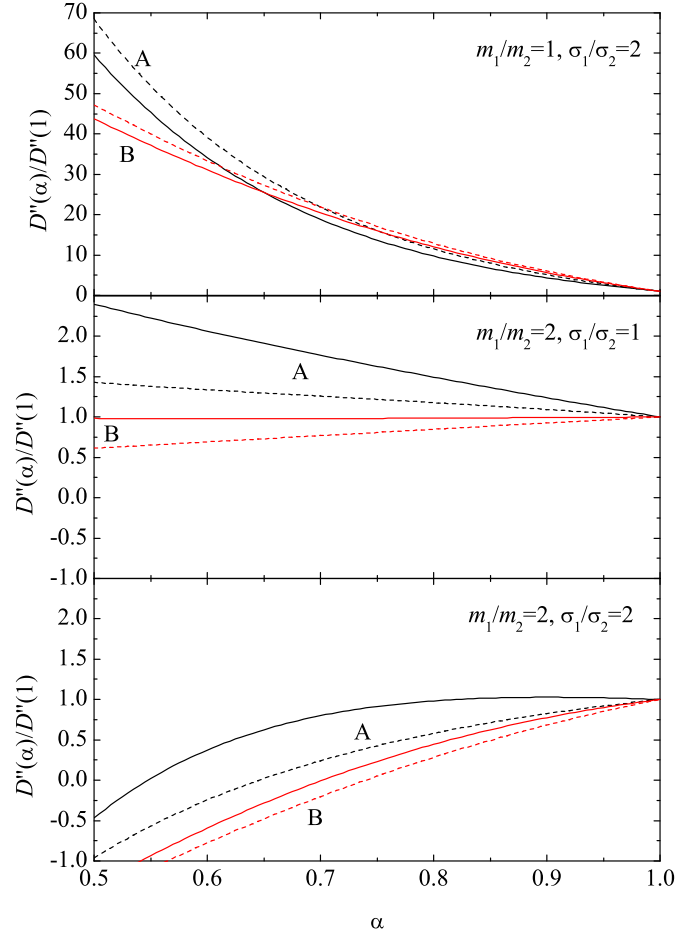}
\caption{Plot of the reduced Dufour coefficient $D''(\alpha)/D''(1)$
as a function of the  coefficient of restitution $\alpha$ for a
three-dimensional equimolar mixture in the cases
$\alpha_{11}=\alpha_{12}=\alpha_{22}=\alpha$ (case A) and
$\alpha_{11}=\alpha$, $\alpha_{12}=(1+\alpha)/2$,
$\alpha_{22}=(3+\alpha)/4$ (case B). The panels correspond, from top
to bottom, to the systems $\{m_1/m_2,\sigma_1/\sigma_2\}=\{1,2\}$,
$\{2,1\}$, and $\{2,2\}$, respectively. The solid lines are the
Boltzmann results, while the dashed lines are the predictions of the
model \protect\eqref{5.5}.}
\label{D2}
\end{figure}

\begin{figure}
\includegraphics[width=.5\columnwidth]{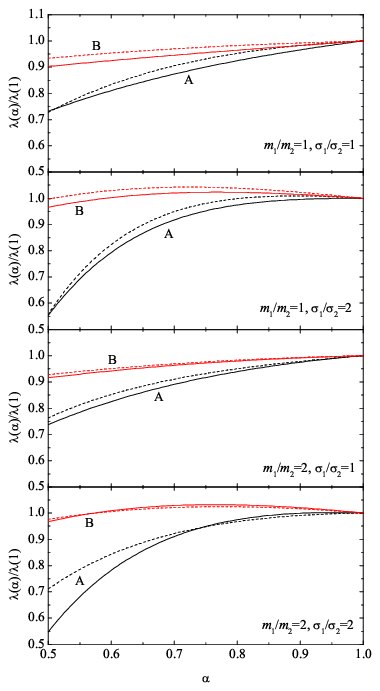}
\caption{Plot of the reduced thermal conductivity coefficient
$\lambda(\alpha)/\lambda(1)$ as a function of the coefficient of
restitution $\alpha$ for a three-dimensional equimolar mixture in
the cases $\alpha_{11}=\alpha_{12}=\alpha_{22}=\alpha$ (case A) and
$\alpha_{11}=\alpha$, $\alpha_{12}=(1+\alpha)/2$,
$\alpha_{22}=(3+\alpha)/4$ (case B). The panels correspond, from top
to bottom, to the systems $\{m_1/m_2,\sigma_1/\sigma_2\}=\{1,1\}$,
$\{1,2\}$, $\{2,1\}$, and $\{2,2\}$, respectively. The solid lines
are the Boltzmann results, while the dashed lines are the
predictions of the model \protect\eqref{5.5}.}
\label{lambda}
\end{figure}

\begin{figure}
\includegraphics[width=.5\columnwidth]{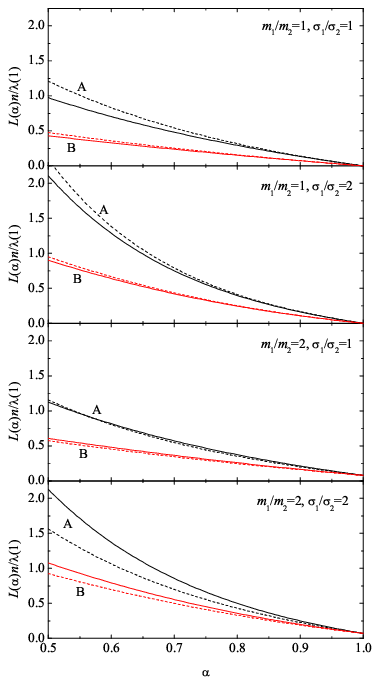}
\caption{Plot of the reduced pressure energy coefficient
$L(\alpha)n/\lambda(1)$ as a function of the coefficient of
restitution $\alpha$ for a three-dimensional equimolar mixture in
the cases $\alpha_{11}=\alpha_{12}=\alpha_{22}=\alpha$ (case A) and
$\alpha_{11}=\alpha$, $\alpha_{12}=(1+\alpha)/2$,
$\alpha_{22}=(3+\alpha)/4$ (case B). The panels correspond, from top
to bottom, to the systems $\{m_1/m_2,\sigma_1/\sigma_2\}=\{1,1\}$,
$\{1,2\}$, $\{2,1\}$, and $\{2,2\}$, respectively. The solid lines
are the Boltzmann results, while the dashed lines are the
predictions of the model \protect\eqref{5.5}.}
\label{L}
\end{figure}

To assess the reliability of  the model described by Eqs.\
\eqref{5.5}, \eqref{n5}, and \eqref{4} in inhomogeneous situations,
in this Section we will consider the corresponding expressions for
the NS transport coefficients of a binary mixture ($N=2$) and will
compare them with the results derived from the original Boltzmann
equation \cite{GD02,GMD06}.

The NS coefficients are defined through the constitutive equations
\cite{GD02}
\begin{equation}
{\bf j}_{1}=- \frac{m_{1}m_{2}n}{\rho } D\nabla x_{1}-\frac{\rho
}{T}D^{\prime }\nabla T-\frac{ \rho }{p}D_{p}\nabla p,\quad{\bf
j}_{2}=-{\bf j}_{1},  \label{6k}
\end{equation}
\begin{equation}
P_{k\ell }=p\delta _{k\ell }-\eta \left( \nabla _{k }u_{\ell
}+\nabla _{\ell }u_{k }-\frac{2}{d}\delta _{k\ell} {\bf \nabla \cdot
u}\right), \label{8}
\end{equation}
\begin{equation}
{\bf q}=-T^{2}D^{\prime \prime }\nabla x_{1}-\lambda \nabla
T-L\nabla p,
\label{7k}
\end{equation}
where $\mathbf{j}_1=m_1 n_1(\mathbf{u}_1 -\mathbf{u})$ is the mass
flux of species 1,  $P_{k\ell}$ is the pressure tensor, and
$\mathbf{q}$ is the heat flux. The transport coefficients in the
constitutive equations are
\begin{equation}
\left(
\begin{array}{c}
D \\
D^{\prime } \\
D_{p} \\
\eta\\
D^{\prime \prime } \\
\lambda  \\
L
\end{array}
\right) =\left(
\begin{array}{c}
\text{diffusion coefficient} \\
\text{thermal diffusion coefficient} \\
\text{pressure diffusion coefficient} \\
\text{shear viscosity}\\
\text{Dufour coefficient} \\
\text{thermal conductivity} \\
\text{pressure energy coefficient}
\end{array}
\right)   \label{9}
\end{equation}
Explicit expressions for the above coefficients are obtained by
solving the inelastic Boltzmann equation by means of the
Chapman--Enskog method. These coefficients are formally given in
terms of the solutions of coupled linear integral equations
involving the linearized Boltzmann collision operators
\begin{equation}
{\cal L}_{1}f_{1}^{(1)}=-
J_{11}^{\inel}[f_{1}^{(0)},f_{1}^{(1)}]-J_{11}^{\inel}[f_{1}^{(1)},f_{1}^{(0)}]-
J_{12}^{\inel}[f_{1}^{(1)},f_{2}^{(0)}] ,
\label{3.12.1}
\end{equation}
\begin{equation}
{\cal M}_{1}f_{2}^{(1)}=-J_{12}^{\inel}[f_{1}^{(0)},f_{2}^{(1)}].
\label{3.12.2}
\end{equation}
The corresponding expressions for the operators ${\cal L}_{2}$ and
${\cal M}_{2}$ can be easily obtained from Eqs.\ (\ref{3.12.1}) and
(\ref{3.12.2}) by just making the changes $1\leftrightarrow 2$. In
the above equations, $f_i^{(0)}$ is the local version of the
distribution function of species $i$ in the HCS \cite{GD99b}.

In the case of the model \eqref{5.5}, one can follow the same formal
steps as in the case of the true Boltzmann equation, except that the
operators $\mathcal{L}_1$ and $\mathcal{M}_1$ are now
\beqa
{\cal L}_{1}f_{1}^{(1)}&=&-\frac{1+\alpha_{11}}{2}\left(
J_{11}^\el[f_{1}^{(0)},f_{1}^{(1)}]+J_{11}^\el[f_{1}^{(1)},f_{1}^{(0)}]\right)-
\frac{1+\alpha_{12}}{2}J_{12}^\el[f_{1}^{(1)},f_{2}^{(0)}]\nn &&
-\frac{\zeta_{11}+\zeta_{12}}{2}\left(\frac{\partial}{\partial {\bf
v}}\cdot {\bf V}f_1^{(1)}- \frac{{\bf
j}_1}{\rho_1}\cdot\frac{\partial}{\partial {\bf v}}f_1^{(0)}\right),
\label{3.12.1b}
\eeqa
\begin{equation}
{\cal
M}_{1}f_{2}^{(1)}=-\frac{1+\alpha_{12}}{2}J_{12}^\el[f_{1}^{(0)},f_{2}^{(1)}].
\label{15}
\end{equation}
In Eq.\ \eqref{3.12.1b}, the drag coefficients \eqref{4} are those
of the HCS, namely
\beq
\label{10k}
\zeta_{ij}=\frac{1}{2}\xi_{ij}\mu_{ji}^2\left(1+\frac{m_i\gamma_j}{m_j
\gamma_i}\right)(1-\alpha_{ij}^2).
\end{equation}

The integral equations defining the transport coefficients are
usually solved by expanding in Sonine polynomials. For practical
purposes, only the leading terms are retained. In addition, we will
use the Maxwellian distribution \eqref{a11} as  a trial function for
$f_i^{(0)}$. The final expressions for the seven transport
coefficients of the mixture are given in Appendix \ref{appB}. A
symbolic code providing the transport properties under arbitrary
values of composition, masses, sizes, and coefficients of
restitution can be downloaded from the website given in Ref.\
\cite{website}.

 Let us compare the predictions of the model \eqref{5.5} for the NS transport coefficients
with the results derived from the original inelastic Boltzmann
equation \cite{GD02,GMD06,GM06}. The three coefficients $D$, $D'$,
and $D_p$ associated with the mass flux are identical in both
descriptions. This is a consequence of the requirement \eqref{n2.7}.
For this reason, these three coefficients will not be shown here.
Their behaviors for several representative cases have been analyzed
elsewhere \cite{GMD06}. The other four coefficients differ in both
descriptions, except trivially in the elastic case
($\alpha_{ij}=1$). Since the parameter space of the problem is
six-dimensional, namely
$\{x_1,m_1/m_2,\sigma_1/\sigma_2,\alpha_{11},\alpha_{12},\alpha_{22}\}$,
it is convenient to choose some specific cases. First, we consider
hard spheres ($d=3$) with a common coefficient of restitution, i.e.,
$\alpha_{11}=\alpha_{12}=\alpha_{22}=\alpha$, (case A) and also with
the choice of the coefficients of restitution $\alpha_{11}=\alpha$,
$\alpha_{12}=(1+\alpha)/2$, $\alpha_{22}=(3+\alpha)/4$ (case B).
Note that in case B one has $\alpha_{11}<\alpha_{12}<\alpha_{22}$.
Consequently, for given values of $x_1$, $m_1/m_2$,
$\sigma_1/\sigma_2$, and $\alpha_{11}=\alpha$, the system A is more
inelastic than the sytem B. Next, we restrict ourselves to equimolar
mixtures ($x_1=\frac{1}{2}$). This reduces the parameter space to
three quantities, namely $\{m_1/m_2,\sigma_1/\sigma_2,\alpha\}$. To
focus on the influence of inelasticity on the transport
coefficients, we fix the values of the mass and size ratios and plot
the transport coefficients as functions of $\alpha$. In addition,
the coefficients are  reduced with respect to their values in the
elastic limit, except in the case of the pressure energy coefficient
$L$, which vanishes for elastic collisions if $m_1=m_2$. In this
case, the plotted quantity is $L(\alpha)n/\lambda(1)$, where
$\lambda(1)$ denotes the elastic value of the thermal conductivity
coefficient. As representative cases we have chosen
$\{m_1/m_2,\sigma_1/\sigma_2\}=\{1,1\}$, $\{1,2\}$, $\{2,1\}$, and
$\{2,2\}$. Note that the system
$\{m_1/m_2,\sigma_1/\sigma_2\}=\{1,1\}$ corresponds to mechanically
equivalent particles in case A.

The shear viscosity is shown in Fig.\ \ref{eta}. We observe that, at
least for the cases analyzed here, the model underestimates the
Boltzmann values. Although the relative discrepancies increase with
dissipation, they are practically insensitive to the mass and size
ratios. As a consequence, the model captures well the influence of
$m_1/m_2$ and $\sigma_1/\sigma_2$ on $\eta$. In particular, it is
interesting to note that the ratio $\eta(\alpha)/\eta(1)$ for the
system $\{m_1/m_2,\sigma_1/\sigma_2\}=\{1,1\}$ is practically
indistinguishable from that of the system
$\{m_1/m_2,\sigma_1/\sigma_2\}=\{2,2\}$, this effect being
reproduced by the model. As expected, the influence of dissipation
is less significant in case B than in case A.

Now we consider the three coefficients associated with the heat
flux. We start with the Dufour coefficient $D''$, which is plotted
in Fig.\ \ref{D2}. The system
$\{m_1/m_2,\sigma_1/\sigma_2\}=\{1,1\}$  is not displayed since
$D''=0$ at any value of $\alpha$ for mechanically equivalent
particles (case A). In case B, however, $D''\neq 0$ for the system
$\{m_1/m_2,\sigma_1/\sigma_2\}=\{1,1\}$ since $\alpha_{11}\neq
\alpha_{12}\neq \alpha_{22}$. We have checked in that case that the
performance of the model is quite good. It is apparent from Fig.\
\ref{D2} that, for a given value of $\alpha$, the ratio
$D''(\alpha)/D''(1)$ has a significant dependence on $m_1/m_2$
and/or $\sigma_1/\sigma_2$. This effect is well accounted for by the
model. It is noteworthy the dramatic influence of inelasticity on
the value of the Dufour coefficient when $m_1=m_2$ and $\sigma_1\neq
\sigma_2$. As shown in the top panel of Fig.\ \ref{D2}, this feature
is accurately predicted by the model.

The thermal conductivity  and the pressure energy coefficients are
displayed in Figs.\ \ref{lambda} and \ref{L}, respectively. In
contrast to $D''$, these coefficients are meaningful in the case of
mechanically equivalent particles. The model performs quite  good a
job, even for strong dissipation, except perhaps for the most
disparate mixture $\{m_1/m_2,\sigma_1/\sigma_2\}=\{2,2\}$ in case A.

In summary, it is fair to say that the model of elastic spheres
subject to a drag force mimics, at least at a semi-quantitative
level, the influence of inelasticity on the transport coefficients
of a binary mixture of inelastic hard spheres. More specifically,
the mass flux coefficients ($D$, $D'$, and $D_p$) are the same in
both systems, while the shear viscosity $\eta$ is underestimated. In
the least favorable case (case A), the discrepancies of the values
of $\eta$ are about 7\% at $\alpha=0.8$ and $13\%$ at $\alpha=0.6$
for the systems studied here. With respect to the heat flux
coefficients ($D''$, $\lambda$, and $L$), the general agreement is,
paradoxically, better than in the case of the shear viscosity. For
instance, the deviations of the thermal conductivity are less than
3\% at $\alpha=0.8$ and 8\% at $\alpha=0.6$ for the systems studied
here in case A. It is also interesting to remark that the
reliability of the transport coefficients predicted by the model is
practically independent of the disparity in mass and size ratio.

\section{Discussion\label{sec4}}
The study of multi-component granular systems is of paramount
importance from a practical point of view, but also at a fundamental
level. In the low density regime, granular gases are well described
by the inelastic Boltzmann equations with constant coefficients of
restitution $\alpha_{ij}\leq 1$. However, the intricacy of the
inelastic Boltzmann collision operator $J_{ij}^\inel$ makes it
difficult to extract explicit results, especially at moderate or
strong dissipation. This motivates the search for models that, while
retaining the basic properties of $J_{ij}^\inel$, provide a simpler
framework for granular mixtures.

In this paper, we have proposed a model of elastic hard spheres
under the action of an external drag force. The role of the force is
to mimic the collisional energy loss in the true gas of inelastic
spheres. In addition, the collision rate for  elastic collisions is
modified by a factor $\beta_{ij}$ with respect to the one for
inelastic collisions. Accordingly, in this model the collision
operator $J_{ij}^\inel$ is replaced by the operator $K_{ij}$ defined
in Eq.\ \eqref{5.5}. While  the dependence of $J_{ij}^\inel$ on
$\alpha_{ij}$ appears in the gain term and through the collision
rules [see Eqs.\ \eqref{2.2} and \eqref{2.3}], such a dependence
appears explicitly in $K_{ij}$ through the parameters $\beta_{ij}$
and $\zeta_{ij}$. These parameters are determined by imposing that
the model operator $K_{ij}$ reproduces the correct collisional
transfer of momentum and energy, Eqs.\ \eqref{n2.7} and
\eqref{n2.7bis}. The former condition gives the simple expression
\eqref{n5} for the parameter $\beta_{ij}$ modifying the collision
rate. The energy condition \eqref{n2.7bis} shows that the drag
coefficient $\zeta_{ij}$ is a functional of the distribution
functions $f_i$ and $f_j$ that is simply proportional to
$1-\alpha_{ij}^2$, Eq.\ \eqref{n7}. Thus, $\zeta_{ij}$ can be
interpreted as the cooling rate of species $i$ due to collisions
with particles of species $j$, as indicated by Eq.\ \eqref{new}.
Since  $\zeta_{ij}$ requires the average $\langle
g_{12}^3\rangle_{ij}$ defined by Eq.\ \eqref{n8}, the Gaussian forms
\eqref{A1} parametrized by the partial velocities and temperatures
are used to estimate $\zeta_{ij}$, yielding Eq.\ \eqref{4}.

By construction, the model leads to the same results for the
temperature ratios in homogeneous states as those obtained from the
inelastic Boltzmann equation in the multi-temperature Maxwellian
approximation \cite{GD99b,BT02}, which is known to compare quite
well with computer simulations \cite{BT02,MG02,DHGD02}. To measure
the performance of the model in inhomogeneous situations, the seven
NS transport coefficients for a binary mixture predicted by the
model have been computed and compared with previous  results from
the inelastic Boltzmann equation \cite{GD02,GMD06,GM06}. The three
transport coefficients associated with the mass flux are identical
in both descriptions in the first Sonine approximation, while the
remaining four coefficients differ, as illustrated in Figs.\
\ref{eta}--\ref{L}. We found that the model captures reasonably well
the dependence of the shear viscosity $\eta$ and the three heat flux
coefficients $D''$, $\lambda$, and $L$ on dissipation. This
agreement is especially remarkable in the case of the thermal
conductivity $\lambda$. The degree of agreement between the model
and the inelastic Boltzmann transport coefficients has been found to
be practically independent of the disparity of mass and/or size,
being similar to the one found for the one-component case
\cite{SA05}. Beyond the NS domain, where the spatial gradients are
not sufficiently small, it is uncertain whether or not the model
predictions are close to the inelastic Boltzmann ones. However, at
least in the uniform shear flow problem, the simulation results in
the one-component case are practically indistinguishable in both
systems \cite{AS05}. We expect that such a good agreement is also
kept for multicomponent systems, given that the discrepancies
between the model and the inelastic Boltzmann transport coefficients
are not significantly affected by the mass and size ratios.

One of the main features of the model \eqref{5.5} is that it allows,
in an approximate way, to ``disentangle'' in $J_{ij}^\inel$ the
purely dissipative effects (represented by the drag-force term) from
those (represented by $\beta_{ij}J_{ij}^\el$) which are essentially
present in the elastic case. Taking advantage of this decomposition,
one can extend to the inelastic case any model kinetic equation
proposed for ordinary multicomponent gases. This is quite an
important issue since, even for elastic collisions, the Boltzmann
equation is too complex to study far from equilibrium situations
\cite{GS03}, for which the NS description fails. For one-component
ordinary gases, the prototype model kinetic equation is the
Bhatnagar--Gross--Krook (BGK) model \cite{BGK54,W54}. Several
extensions of this model to the case of one-component inelastic hard
spheres have been proposed \cite{BMD96,BDS99,DBZ04}. Furthermore,
some kinetic models for ordinary gas mixtures inspired on the BGK
model can be found in the literature \cite{GS03,GK56,GSB89,AAP02}.
The common structure of the latter models is
\beq
J_{ij}^\el[\mathbf{v}|f_i,f_j]\to
-\nu_{ij}\left[f_i(\mathbf{v})-f_{ij}(\mathbf{v})\right],
\label{4.10}
\eeq
where $\nu_{ij}$ is a velocity-independent effective collision
frequency of a particle of species $i$ with particles of species $j$
and $f_{ij}(\mathbf{v})$ is a reference distribution function whose
velocity dependence is explicit and that involves a number of
parameters to be determined by imposing that Eq.\ \eqref{4.10}
retains the main physical properties of the Boltzmann operator
$J_{ij}^\el$. Using the mapping \eqref{5.5}, any kinetic model of
the form \eqref{4.10} can be easily extended to the case of
inelastic collisions as
\beq
J_{ij}^\inel[\mathbf{v}|f_i,f_j]\to
-\beta_{ij}\nu_{ij}\left[f_i(\mathbf{v})-f_{ij}(\mathbf{v})\right]+\frac{\zeta_{ij}}{2}\frac{\partial
}{\partial {\bf v}}\cdot \left[\left( {\bf v
}-\mathbf{u}_{i}\right)f_i(\mathbf{v})\right],
\label{4.11}
\eeq
where $\beta_{ij}$ and $\zeta_{ij}$ are given by Eqs.\ \eqref{n5}
and \eqref{4}, respectively. A specific choice for $f_{ij}$ based on
the Gross--Krook kinetic model \cite{GK56} is worked out in Appendix
\ref{appC}. Several applications of the kinetic model \eqref{4.11}
to non-Newtonian flows are under way and will be published
elsewhere.

\acknowledgments

This research  has been supported by the Ministerio de Educaci\'on y
Ciencia (Spain) through Programa Juan de la Cierva (F.V.R.) and
Grant No.\ FIS2004-01399, partially financed by FEDER funds.

\appendix

\section{Evaluation of $\langle g_{12}^3\rangle_{ij}$,
$\langle g_{12}\mathbf{g}_{12}\rangle_{ij}$, and $\langle g_{12}\mathbf{g}_{12}\cdot
\mathbf{G}_{12}\rangle_{ij}$ \label{appA}}

The aim of this Appendix is to evaluate the averages $\langle g_{12}^3\rangle_{ij}$, $\langle
g_{12}\mathbf{g}_{12}\rangle_{ij}$, and $\langle g_{12}\mathbf{g}_{12}\cdot \mathbf{G}_{12}\rangle_{ij}$ by
assuming the Gaussian forms (\ref{A1}).

Let us start with the evaluation of $\langle g_{12}^3\rangle_{ij}$.
Inserting Eq.\ \eqref{A1} into Eq.\ \eqref{n8}, one gets
\beq
\langle
g_{12}^3\rangle_{ij}=\left(\frac{m_i}{2\pi\widetilde{T}_i}\right)^{d/2}
\left(\frac{m_j}{2\pi\widetilde{T}_j}\right)^{d/2}
\int \dd\mathbf{w}_1 \int \dd\mathbf{w}_2
\exp\left(-\frac{m_iw_1^2}{2\widetilde{T}_i}-\frac{m_jw_2^2}{2\widetilde{T}_j}
\right)|\mathbf{w}_{12}+\mathbf{u}_i-\mathbf{u}_j|^3,
\label{A2}
\eeq
where $\mathbf{w}_1=\mathbf{v}_1-\mathbf{u}_i$,
$\mathbf{w}_2=\mathbf{v}_2-\mathbf{u}_j$, and
$\mathbf{w}_{12}=\mathbf{w}_1-\mathbf{w}_2$. Let us also introduce
the variable
\beq
\mathbf{W}_{12}=\left(\frac{m_i}{\widetilde{T}_i}+\frac{m_j}{\widetilde{T}_j}\right)^{-1}
\left(\frac{m_i}{\widetilde{T}_i}\mathbf{w}_1
+\frac{m_j}{\widetilde{T}_j}\mathbf{w}_2\right),
\label{A3}
\eeq
so that $\dd\mathbf{w}_1
\dd\mathbf{w}_2=\dd\mathbf{w}_{12}\dd\mathbf{W}_{12}$ and
\beq
\frac{m_iw_1^2}{\widetilde{T}_i}+\frac{m_jw_2^2}{\widetilde{T}_j}=\left(\frac{m_i}
{\widetilde{T}_i}+\frac{m_j}{\widetilde{T}_j}\right)W_{12}^2
+\left(\frac{\widetilde{T}_i}{m_i}+\frac{\widetilde{T}_j}{m_j}\right)^{-1}w_{12}^2.
\label{A4}
\eeq
Thus, after making the changes $(\mathbf{v}_1,\mathbf{v}_2)\to
(\mathbf{w}_{12}, \mathbf{W}_{12})$ and integrating over
$\mathbf{W}_{12}$, Eq.\ \eqref{A2} becomes
\beqa
\langle g_{12}^3\rangle_{ij}&=&\pi^{-d/2}
\left(\frac{2\widetilde{T}_i}{m_i}+\frac{2\widetilde{T}_j}{m_j}\right)^{-d/2}
\int \dd\mathbf{w}_{12}
\exp\left[-\left(\frac{2\widetilde{T}_i}{m_i}+\frac{2\widetilde{T}_j}{m_j}\right)^{-1}w_{12}^2\right]|
\mathbf{w}_{12}+\mathbf{u}_i-\mathbf{u}_j|^3\nn
&=&
\left(\frac{2\widetilde{T}_i}{m_i}+\frac{2\widetilde{T}_j}{m_j}\right)^{3/2}
\pi^{-d/2}\int \dd\mathbf{c}\,
e^{-c^2}|\mathbf{c}+\mathbf{u}_{ij}^*|^3,
\label{A5}
\eeqa
where in the second step we have made the change of variable
$\mathbf{c}=(2\TT_i/m_i+2\TT_j/m_j)^{-1/2}\mathbf{w}_{12}$ and have
defined
$\mathbf{u}_{ij}^*=(2\TT_i/m_i+2\TT_j/m_j)^{-1/2}(\mathbf{u}_{i}-\mathbf{u}_{j})$.
The Gaussian integral in Eq.\ \eqref{A5} cannot be evaluated
analytically for arbitrary $\mathbf{u}_{ij}^*$. Here we will
evaluate it by neglecting terms of fourth and higher order in
$\mathbf{u}_{ij}^*$. In that case, one has
\beq
|\mathbf{c}+\mathbf{u}_{ij}^*|^3=c^3+3c \mathbf{c}\cdot
\mathbf{u}_{ij}^*+\frac{3}{2}c {{u}_{ij}^*}^2+\frac{3}{2}
c^{-1}\left(\mathbf{c}\cdot \mathbf{u}_{ij}^*\right)^2+\cdots,
\label{A6}
\eeq
where we have taken into account that the third order terms do not
contribute to the integral. As a consequence,
\beqa
\pi^{-d/2}\int \dd\mathbf{c}\,
e^{-c^2}|\mathbf{c}+\mathbf{u}_{ij}^*|^3&\to&\pi^{-d/2}\int
\dd\mathbf{c}\, e^{-c^2}\left(c^3+\frac{3}{2}\frac{d+1}{d}c
{{u}_{ij}^*}^2\right)\nn
&=&\frac{\Gamma((d+1)/2)}{\Gamma(d/2)}\frac{d+1}{2}\left(1+\frac{3}{d}{u_{ij}^*}^2\right).
\label{A7}
\eeqa
Insertion of \eqref{A7} into \eqref{A5} yields
\beq
\langle
g_{12}^3\rangle_{ij}=(d+1)\frac{\Gamma((d+1)/2)}{\Gamma(d/2)}
\left(\frac{2\widetilde{T}_i}{m_i}+\frac{2\widetilde{T}_j}{m_j}\right)^{1/2}
\left[\frac{\widetilde{T}_i}{m_i}+\frac{\widetilde{T}_j}{m_j}+\frac{3}{2d}
\left(\mathbf{u}_i-\mathbf{u}_j\right)^2\right].
\label{n11}
\eeq

Following similar steps, it is straightforward to get
\beq
\langle g_{12}\mathbf{g}_{12}\rangle_{ij}=
\left(\frac{2\widetilde{T}_i}{m_i}+\frac{2\widetilde{T}_j}{m_j}\right)
\pi^{-d/2}\int \dd\mathbf{c}\,
e^{-c^2}|\mathbf{c}+\mathbf{u}_{ij}^*|(\mathbf{c}+\mathbf{u}_{ij}^*).
\label{A8}
\eeq
Next, one has
\beq
|\mathbf{c}+\mathbf{u}_{ij}^*|(\mathbf{c}+\mathbf{u}_{ij}^*)=c\left(\mathbf{u}_{ij}^*+
c^{-2}\mathbf{u}_{ij}^*\cdot
\mathbf{c}\mathbf{c}+\cdots\right),
\label{A9}
\eeq
where the ellipsis denote terms which are odd in $\mathbf{c}$ or are
of at least of third order in $\mathbf{u}_{ij}^*$. Thus,
\beqa
\pi^{-d/2}\int \dd\mathbf{c}\,
e^{-c^2}|\mathbf{c}+\mathbf{u}_{ij}^*|(\mathbf{c}+\mathbf{u}_{ij}^*)&\to&\mathbf{u}_{ij}^*
\frac{d+1}{d}\pi^{-d/2}\int
\dd\mathbf{c}\, e^{-c^2}c\nn
&=&\frac{d+1}{d}\frac{\Gamma((d+1)/2)}{\Gamma(d/2)}\mathbf{u}_{ij}^*.
\label{A10}
\eeqa
{}From Eqs.\ \eqref{A8} and \eqref{A9} it follows that
\beq
\langle g_{12}\mathbf{g}_{12}\rangle_{ij}=
\frac{d+1}{d}\frac{\Gamma((d+1)/2)}{\Gamma(d/2)}\left(\frac{2\widetilde{T}_i}{m_i}+
\frac{2\widetilde{T}_j}{m_j}\right)^{1/2}
(\mathbf{u}_{i}-\mathbf{u}_{j}).
\label{4.3}
\eeq

Finally, we consider the evaluation of $\langle
g_{12}\mathbf{g}_{12}\cdot \mathbf{G}_{12}\rangle_{ij}$. Taking into
account Eq.\ \eqref{A3} and the definition of $\mathbf{G}_{12}$ [see
below Eq.\ \eqref{n6}], it is easy to get the relationship
\beq
\mathbf{G}_{12}=\mathbf{W}_{12}+\left(\frac{m_i}{\TT_i}+\frac{m_j}{\TT_j}\right)^{-1}\left[\left(\frac{m_i}{\TT_i}\mathbf{u}_i+
\frac{m_j}{\TT_j}\mathbf{u}_j\right)-\frac{m_im_j}{m_i+m_j}\left(\TT_i^{-1}-\TT_j^{-1}\right)
\mathbf{g}_{12}\right].
\label{A11}
\eeq
As a consequence,
\beq
\langle g_{12}\mathbf{g}_{12}\cdot \mathbf{G}_{12}\rangle_{ij}=
\left(\frac{m_i}{\TT_i}+\frac{m_j}{\TT_j}\right)^{-1}\left[\langle
g_{12}\mathbf{g}_{12}\rangle_{ij}\cdot\left(\frac{m_i}{\TT_i}\mathbf{u}_i+\frac{m_j}{\TT_j}
\mathbf{u}_j\right)-\frac{m_im_j}{m_i+m_j}\left(\TT_i^{-1}-\TT_j^{-1}\right)
\langle{g}_{12}^3\rangle_{ij}\right],
\label{A12}
\eeq
where, by symmetry, $\langle g_{12}\mathbf{g}_{12}\cdot
\mathbf{W}_{12}\rangle_{ij}=0$ when evaluated with the distributions
\eqref{A1}. Making use of Eqs.\ \eqref{n11} and \eqref{4.3} one gets
\beqa
\langle g_{12}\mathbf{g}_{12}\cdot
\mathbf{G}_{12}\rangle_{ij}&=&\frac{d+1}{d}\frac{\Gamma((d+1)/2)}{\Gamma(d/2)}
\left(\frac{2\widetilde{T}_i}{m_i}+\frac{2\widetilde{T}_j}{m_j}\right)^{1/2}
\left(\frac{m_i}{\TT_i}+\frac{m_j}{\TT_j}\right)^{-1}\nn
&&\times
\left\{\left(\frac{m_i}{\TT_i}\mathbf{u}_i+\frac{m_j}{\TT_j}\mathbf{u}_j\right)
\cdot\left(\mathbf{u}_i-\mathbf{u}_j\right)\right.\nn
&&\left. -d\frac{m_im_j}{m_i+m_j}\left(\TT_i^{-1}-\TT_j^{-1}\right)
\left[\frac{\widetilde{T}_i}{m_i}+\frac{\widetilde{T}_j}{m_j}+\frac{3}{2d}
\left(\mathbf{u}_i-\mathbf{u}_j\right)^2\right]\right\}.
\label{4.5}
\eeqa

\section{Explicit expressions for the transport coefficients
\label{appB}}
In this Appendix, we take advantage of the results derived from the
original Boltzmann equation \cite{GD02,GM06} to determine the
expressions corresponding to the model \eqref{5.5}, by using the
forms \eqref{3.12.1b} and \eqref{15} for the operators
$\mathcal{L}_i$ and $\mathcal{M}_i$.

\subsection{Mass flux}
In the case of the transport coefficients associated with the mass
flux, the results are \cite{GD02,GM06}
\begin{equation}
D=\frac{\rho }{m_{1}m_{2}n}\left( \nu -\frac{1}{2}\zeta\right) ^{-1}
\left[ p\left( \frac{\partial }{\partial
x_{1}}x_{1}\gamma_{1}\right) _{p,T}+\rho \left( \frac{\partial
\zeta}{\partial x_{1}}\right) _{p,T}\left( D_{p}+D^{\prime }\right)
\right] ,  \label{a13}
\end{equation}
\begin{equation}
D_{p}=\frac{n_{1}T}{\rho }\left( \gamma_1-\frac{m_{1}n}{\rho
}\right) \left( \nu -\frac{3}{2}\zeta+\frac{\zeta ^{2}}{2\nu
}\right) ^{-1},
\label{a14}
\end{equation}
\begin{equation}
D^{\prime }=-\frac{\zeta}{2\nu }D_{p}.  \label{a15}
\end{equation}
Here, $\zeta=\zeta_1=\zeta_2$ is the cooling rate of the mixture in
the HCS [see Eq.\ \eqref{n15}] and the collision frequency $\nu$ is
given by
\begin{equation}
\nu =\frac{m_{1}}{dn_{1}T\gamma_{1}}\int \dd{\bf v}{\bf V}\cdot
\left(\mathcal{L}_1 f_{1,M}{\bf
V}-\frac{x_1\gamma_1}{x_2\gamma_2}\mathcal{M}_1 f_{2,M}{\bf
V}\right) .
\label{a10}
\end{equation}
 The result is
\begin{equation}
\label{3.16}
\nu=\nu_0\frac{2\pi^{(d-1)/2}}{d\Gamma\left(\frac{d}{2}\right)}(1+\alpha_{12})
\left(\frac{\theta_1+\theta_2}{\theta_1\theta_2}\right)^{1/2}\left(x_2\mu_{21}+
x_1\mu_{12}\right),
\end{equation}
where $\nu_0\equiv n\sigma_{12}^{d-1}\sqrt{2T(m_1+m_2)/m_1m_2}$,
$\theta_1\equiv (\mu_{21}\gamma_1)^{-1}$,  $\theta_2\equiv
(\mu_{12}\gamma_2)^{-1}$, and we recall that $\gamma_i\equiv T_i/T$.
The expressions of the transport coefficients $D$, $D_p$, and $D'$
are exactly the same as those obtained from the true Boltzmann
equation.

\subsection{Pressure tensor}
The shear viscosity coefficient $\eta$ can be written as
\begin{equation}
\label{3.17}
\eta=\frac{p}{\nu_0}\left(x_1\gamma_1^2\eta_1^*+x_2\gamma_2^2\eta_2^*\right),
\end{equation}
where the expression of the (dimensionless) partial contribution
$\eta_1^*$ is
\begin{equation}
\label{3.18}
\eta_1^*=2\frac{\gamma_2(2\tau_{22}-\zeta^{*})-2\gamma_1\tau_{12}}
{\gamma_1\gamma_2[\zeta^*-2\zeta^{*}
(\tau_{11}+\tau_{22})+4(\tau_{11}\tau_{22}-\tau_{12}\tau_{21})]}.
\end{equation}
Here, $\zeta^*\equiv \zeta/\nu_0$ and we have introduced the
(reduced) collision frequencies
\begin{equation}
\label{b13}
\tau_{11}=\frac{1}{(d-1)(d+2)}\frac{1}{n_1 T^2\gamma_1^2\nu_0}\int
\dd{\bf v} R_{1,k\ell} {\cal L}_1\left(f_{1,M}R_{1,k\ell}\right),
\end{equation}
\begin{equation}
\label{b14}
\tau_{12}=\frac{1}{(d-1)(d+2)}\frac{1}{n_1T^2\gamma_1^2\nu_0}\int
\dd{\bf v} R_{1,k\ell} {\cal M}_1\left(f_{2,M}R_{2,k\ell}\right),
\end{equation}
where
\begin{equation}
\label{b9}
R_{1,k\ell}({\bf V})=m_1\left( V_{k}V_{\ell}-
\frac{1}{d}V^2\delta_{k\ell}\right).
\end{equation}
A similar expression can be obtained for $\eta_2^*$ by just making
the changes $1 \leftrightarrow 2$. Taking advantage of the results
derived for the $d$-dimensional Boltzmann equation \cite{GM06}, one
gets for the model the expressions
\begin{widetext}
\begin{eqnarray}
\label{3.19}
\tau_{11}&=&\frac{4\pi^{(d-1)/2}}{d(d+2)\Gamma\left(\frac{d}{2}\right)}\left\{
x_1\left(\frac{\sigma_{1}}{\sigma_{12}}\right)^{d-1}(2\theta_1)^{-1/2}d
(1+\alpha_{11})\right.\nonumber\\
& &+x_2 \mu_{21}(1+\alpha_{12}) \theta_1^{3/2}\theta_2^{-1/2} \left[
(d+3)(\mu_{12}\theta_2-\mu_{21}\theta_1)\theta_1^{-2}(\theta_1+\theta_2)^{-1/2}\right.\nonumber\\
& & \left.\left.+d\mu_{21}\theta_1^{-2}(\theta_1+\theta_2)^{1/2}
+\frac{2d(d+1)-4}{2(d-1)}\theta_1^{-1}(\theta_1+\theta_2)^{-1/2}\right]\right\}+\zeta_{11}^*+\zeta_{12}^*,
\end{eqnarray}
\begin{eqnarray}
\label{3.20}
\tau_{12}&=&\frac{4\pi^{(d-1)/2}}{d(d+2)\Gamma\left(\frac{d}{2}\right)}
x_2\frac{\mu_{21}^2}{\mu_{12}}\theta_1^{3/2}\theta_2^{-1/2}
(1+\alpha_{12})\nonumber\\
& \times&\left[
(d+3)(\mu_{12}\theta_2-\mu_{21}\theta_1)\theta_2^{-2}(\theta_1+\theta_2)^{-1/2}\right.\nonumber\\
& & \left.+d\mu_{21}\theta_2^{-2}(\theta_1+\theta_2)^{1/2}
-\frac{2d(d+1)-4}{2(d-1)}\theta_2^{-1}(\theta_1+\theta_2)^{-1/2}\right],
\end{eqnarray}
\end{widetext}
where $\zeta_{ij}^*\equiv \zeta_{ij}/\nu_0$ and $\zeta_{íj}$ is given by Eq.\ (\ref{10k}).

\subsection{Heat flux}
 The transport coefficients
appearing in the heat flux ($D''$, $L$, and $\lambda$) can be
written as \cite{GD02,GM06}
\begin{equation}
D^{\prime \prime }=-\frac{d+2}{2}\frac{n}{(m_{1}+m_{2})\nu_0}\left[
\frac{ x_{1}\gamma _{1}^{3}}{\mu _{12}}d_{1}^*+\frac{x_{2}\gamma
_{2}^{3}}{\mu _{21}}d_{2}^*-\left( \frac{\gamma _{1}}{\mu
_{12}}-\frac{\gamma _{2}}{\mu _{21}}\right) D^*\right] ,
\label{4.7}
\end{equation}
\begin{equation}
L=-\frac{d+2}{2}\frac{T}{(m_{1}+m_{2})\nu_0}\left[ \frac{x_{1}\gamma
_{1}^{3}}{\mu _{12}}\ell _{1}^*+\frac{x_{2}\gamma _{2}^{3}}{\mu
_{21}}\ell _{2}^*-\left( \frac{\gamma _{1}}{\mu _{12}}-\frac{\gamma
_{2}}{\mu _{21}} \right) D_{p}^*\right] ,
\label{4.8}
\end{equation}
\begin{equation}
\lambda =-\frac{d+2}{2}\frac{nT}{(m_{1}+m_{2})\nu _{0}}\left[ \frac{
x_{1}\gamma _{1}^{3}}{\mu _{12}}\lambda _{1}^*+\frac{x_{2}\gamma
_{2}^{3}}{\mu_{21}}\lambda _{2}^*-\left( \frac{\gamma _{1}}{\mu
_{12}}-\frac{\gamma _{2}}{ \mu _{21}}\right) D^{\prime *}\right] ,
\label{4.9}
\end{equation}
where
\begin{equation}
D=\frac{\rho T}{m_{1}m_{2}\nu _{0}}D^{\ast },\quad
D_{p}=\frac{nT}{\rho \nu _{0}}D_{p}^{\ast },\quad D^{\prime
}=\frac{nT}{\rho \nu _{0}}D^{\prime *},  \label{3.n1}
\end{equation}
the coefficients $D$, $D_{p}$, and
 $D^{\prime}$ being given by
 Eqs.\ (\ref{a13})--(\ref{a15}), respectively. The expressions
 of the (dimensionless)  coefficients $d_{i}^{*}$, $\ell
_{i}^*$, and $\lambda _{i}^*$ are
\begin{eqnarray}
\label{4.n2}
d_1^{*}&=&\frac{1}{\Delta}\Big\{2\left[2
\nuk_{12}Y_2-Y_1(2\nuk_{22}-3\zeta^*)\right]\left[\nuk_{12}\nuk_{21}-\nuk_{11}\nuk_{22}
+2(\nuk_{11}+\nuk_{22})\zeta^*-4\zeta^{*2}\right]\nonumber\\
& &+2\left( \frac{\partial \zeta ^{\ast }}{\partial x_{1}}\right)
_{p,T}(Y_3+Y_5)\left[2\nuk_{12}\nuk_{21}+2\nuk_{22}^2-\zeta^*(
7\nuk_{22}-6\zeta^{*})\right]\nonumber\\
& & -2\nuk_{12}\left( \frac{\partial \zeta ^{\ast }}{\partial
x_{1}}\right)_{p,T}(Y_4+Y_6)\left(2\nuk_{11}+2\nuk_{22}-7\zeta^*\right)\Big\},
\end{eqnarray}
\begin{eqnarray}
\label{4.n3} \ell_1^*&=&\frac{1}{\Delta}\left\{-2Y_3\left[2
(\nuk_{12}\nuk_{21}-\nuk_{11}\nuk_{22})\nuk_{22}+\zeta^*(7\nuk_{11}\nuk_{22}-5\nuk_{12}\nuk_{21}+2\nuk_{22}^2
-6\nuk_{11}\zeta^*-7\nuk_{22}\zeta^*+6\zeta^{*2})\right]\right.\nonumber\\
&&
+2Y_4\nuk_{12}\left[2\nuk_{12}\nuk_{21}-2\nuk_{11}\nuk_{22}+2\zeta^*(\nuk_{11}+\nuk_{22})
-\zeta^{*2}\right]\nonumber\\
& &
\left.+2Y_5\zeta^*\left[2\nuk_{12}\nuk_{21}+\nuk_{22}(2\nuk_{22}-7\zeta^*)+6\zeta^{*2}\right]
-2\nuk_{12}\zeta^*Y_6\left[2(\nuk_{11}+\nuk_{22})-7\zeta^*\right]
\right\},
\end{eqnarray}
\begin{eqnarray}
\label{4.n4}
\lambda_1^*&=&\frac{1}{\Delta}\left\{-Y_3\zeta^*\left[2
\nuk_{12}\nuk_{21}+\nuk_{22}(2\nuk_{22}-7\zeta^*)+6\zeta^{*2}\right]+\nuk_{12}\zeta^*
Y_4\left[2(\nuk_{11}+\nuk_{22})-7\zeta^*\right] \right. \nonumber\\
&
&-Y_5\left[4\nuk_{12}\nuk_{21}(\nuk_{22}-\zeta^*)+2\nuk_{22}^2(5\zeta^*-2\nuk_{11})+2\nuk_{11}
(7\nuk_{22}\zeta^*-6\zeta^{*2})+5\zeta^{*2}(6\zeta^*-7\nuk_{22})\right]\nonumber\\
& & \left.
+\nuk_{12}Y_6\left[4\nuk_{12}\nuk_{21}+2\nuk_{11}(5\zeta^*-2\nuk_{22})+\zeta^*(10\nuk_{22}-
23\zeta^*)\right]\right\}.
\end{eqnarray}
In the above equations, we have introduced the quantities
\begin{equation}
\label{4.n5}
\Delta\equiv\left[4(\nuk_{12}\nuk_{21}-\nuk_{11}\nuk_{22})+6\zeta^*(\nuk_{11}+\nuk_{22})-9\zeta^{*2}\right]
\left[\nuk_{12}\nuk_{21}-\nuk_{11}\nuk_{22}+2\zeta^*(\nuk_{11}+\nuk_{22})-4\zeta^{*2}\right],
\end{equation}
\begin{equation}
\label{4.14}
 Y_1= \frac{D^*}{
x_1\gamma_1^2}\left(\omega_{12}-\zeta^{*}\right)-\frac{1}{
\gamma_1^2} \left(\frac{\partial \gamma_1}{\partial x_1}
\right)_{p,T},\quad Y_2= -\frac{D^*}{
x_2\gamma_2^2}\left(\omega_{21}-\zeta^{*}\right)-\frac{1}{
\gamma_2^2} \left(\frac{\partial \gamma_2}{\partial x_1}
\right)_{p,T},
\end{equation}
\begin{equation}
\label{4.15k}
Y_3=\frac{
D_p^*}{x_1\gamma_1^2}\left(\omega_{12}-\zeta^{*}\right),\quad
Y_4=-\frac{ D_p^*}{x_2\gamma_2^2}\left(\omega_{21}-\zeta^{*}\right),
\end{equation}
\begin{equation}
\label{4.16k}
Y_5= -\frac{1}{\gamma_1}+\frac{D^{\prime\ast}}{
x_1\gamma_1^2}\left(\omega_{12}-\zeta^{*}\right),\quad Y_6=
-\frac{1}{\gamma_2}-\frac{D^{\prime\ast}}{
x_2\gamma_2^2}\left(\omega_{21}-\zeta^{*}\right),
\end{equation}
\begin{equation}
\label{4.17b}
\nuk_{11}=\frac{2}{d(d+2)}\frac{m_1}{n_1T^3\gamma_1^3\nu_0}\int
\dd{\bf v} {\bf S}_1 \cdot {\cal L}_1\left(f_{1,M}{\bf S}_1\right),
\end{equation}
\begin{equation}
\label{4.18b}
\nuk_{12}=\frac{2}{d(d+2)}\frac{m_1}{n_1T^3\gamma_1^3\nu_0}\int
\dd{\bf v} {\bf S}_1 \cdot {\cal M}_1\left(f_{2,M}{\bf S}_2\right),
\end{equation}
\begin{equation}
\label{4.19b}
\omega_{12}=\frac{2}{d(d+2)}\frac{m_1}{n_1T^2\gamma_1^2\nu_0}
\left[\int \dd{\bf v} {\bf S}_1\cdot {\cal L}_1(f_{1,M}{\bf
V}_1)-\frac{x_1\gamma_1}{x_2\gamma_2} \int \dd{\bf v} {\bf S}_1\cdot
{\cal M}_1(f_{2,M}{\bf V}_2)\right],
\end {equation}
where
\beq
{\bf S}_i({\bf
V})=\left(\frac{1}{2}m_iV^2-\frac{d+2}{2}T\gamma_i\right){\bf V}.
\eeq

The explicit expressions for the coefficients $\nuk_{ij}$ and
$\omega_{ij}$ are
\begin{eqnarray}
\label{c18}
\nuk_{11}&=&\frac{\pi^{(d-1)/2}}{\Gamma\left(\frac{d}{2}\right)}
\frac{4}{d(d+2)}\left(\frac{\sigma_1}{\sigma_{12}}\right)^{d-1} x_1
(2\theta_1)^{-1/2}
(1+\alpha_{11})\left(d-1\right)\nonumber\\
& & +\frac{\pi^{(d-1)/2}}{\Gamma\left(\frac{d}{2}\right)}
\frac{1}{d(d+2)}x_2\mu_{21}(1+\alpha_{12})\left(\frac{\theta_1}
{\theta_2(\theta_1+\theta_2)}\right)^{3/2}\left[E-(d+2)\frac{\theta_1+\theta_2}{\theta_1}A\right]
+\frac{3}{2}\left(\zeta_{11}^*+\zeta_{12}^*\right),
\end{eqnarray}
\begin{equation}
\label{c19}
\nuk_{12}=-\frac{\pi^{(d-1)/2}}{\Gamma\left(\frac{d}{2}\right)}
\frac{1}{d(d+2)}x_2\frac{\mu_{21}^2}{\mu_{12}}(1+\alpha_{12})\left(\frac{\theta_1}
{\theta_2(\theta_1+\theta_2)}\right)^{3/2}\left[F+(d+2)\frac{\theta_1+\theta_2}{\theta_2}B\right].
\end{equation}
\beq
\label{c17}
\omega_{12}=\frac{\pi^{(d-1)/2}}
{\Gamma\left(\frac{d}{2}\right)}\frac{2}{d(d+2)}
x_1\mu_{21}(1+\alpha_{12})(\theta_1+\theta_2)^{-1/2}\theta_1^{1/2}
\theta_2^{-3/2}\left(\frac{x_2}{x_1}A-\frac{\gamma_1}{\gamma_2}
B\right)+\zeta_{11}^*+\zeta_{12}^*,
\eeq
where
\begin{eqnarray}
\label{c20}
A&=&(d+2)(2\phi_{12}+\theta_2)+4\mu_{21}(\theta_1+\theta_2)(d-1)\phi_{12}\theta_1^{-1}\nonumber\\
& &
+3(d+3)\phi_{12}^2\theta_1^{-1}+\mu_{21}^2(d+3)\theta_1^{-1}(\theta_1+\theta_2)^2-
(d+2)\theta_2\theta_1^{-1}(\theta_1+\theta_2),
\end{eqnarray}
\begin{eqnarray}
\label{c21}
B&=&
(d+2)(2\phi_{12}-\theta_1)-4\mu_{21}(\theta_1+\theta_2)(d-1)\phi_{12}\theta_2^{-1}\nonumber\\
& &
-3(d+3)\phi_{12}^2\theta_2^{-1}-\mu_{21}^2(d+3)\theta_2^{-1}(\theta_1+\theta_2)^2+
(d+2)(\theta_1+\theta_2),
\end{eqnarray}
\begin{eqnarray}
\label{c22}
E&=&
 \mu_{21}^2\theta_1^{-2}(\theta_1+\theta_2)^2
(d+3)
\left[(d+2)\theta_1+(d+5)\theta_2\right]\nonumber\\
& & +4(d-1)\mu_{21}(\theta_1+\theta_2)
\left\{\phi_{12}\theta_1^{-2}[(d+2)\theta_1+(d+5)\theta_2]+2\theta_2\theta_1^{-1}\right\}
\nonumber\\
& & +3(d+3)\phi_{12}^2\theta_1^{-2}[(d+2)\theta_1+(d+5)\theta_2]+
2\phi_{12}\theta_1^{-1}[(d+2)^2\theta_1+(24+11d+d^2)\theta_2]
\nonumber\\
& & +(d+2)\theta_2\theta_1^{-1}
[(d+8)\theta_1+(d+3)\theta_2]-(d+2)(\theta_1+\theta_2)\theta_1^{-2}\theta_2
[(d+2)\theta_1+(d+3)\theta_2],\nn
\end{eqnarray}
\begin{eqnarray}
\label{c23}
F&=&
 \mu_{21}^2\theta_2^{-2}(\theta_1+\theta_2)^2
(d+3)
\left[(d+5)\theta_1+(d+2)\theta_2\right]\nonumber\\
& & +4(d-1)\mu_{21}(\theta_1+\theta_2)
\left\{\phi_{12}\theta_2^{-2}[(d+5)\theta_1+(d+2)\theta_2]-2\theta_1\theta_2^{-1}\right\}
\nonumber\\
& & +3(d+3)\phi_{12}^2\theta_2^{-2}[(d+5)\theta_1+(d+2)\theta_2]-
2\phi_{12}\theta_2^{-1}[(24+11d+d^2)\theta_1+(d+2)^2\theta_2]
\nonumber\\
& & +(d+2)\theta_1\theta_2^{-1}
[(d+3)\theta_1+(d+8)\theta_2]-(d+2)(\theta_1+\theta_2)\theta_2^{-1}
[(d+3)\theta_1+(d+2)\theta_2].\nn
\end{eqnarray}
Here, $\phi_{12}=\mu_{12}\theta_2-\mu_{21}\theta_1$. The expressions
for $\chi_{22}$, $\chi_{21}$, and $\omega_{21}$ can be easily
obtained from Eqs.\ (\ref{c18})--(\ref{c17}) by exchanging
$1\leftrightarrow 2$.

\section{A model kinetic equation for granular mixtures\label{appC}}
This Appendix addresses the construction of the kinetic model
\eqref{4.11} for a specific choice of the reference distribution
function $f_{ij}$.

The most natural way of extending the BGK model to (elastic) mixtures is to assume a Gaussian form for the
reference distribution $f_{ij}$, i.e.,
\begin{equation}
 f_{ij}(\mathbf{v})=n_i\left(\frac{m_i}{2\pi
T_{ij}}\right)^{d/2} \exp\left[-\frac{m_i}{2T_{ij}}\left(\mathbf{v}-\mathbf{u}_{ij}\right)^2\right],
\label{4.12} \end{equation}
where $\mathbf{u}_{ij}$ and $T_{ij}$ are parameters to be determined. This is the
form of the model proposed by Gross and Krook \cite{GK56}, which has been widely used in the literature. The
usual criteria to determine the unknowns $\nu_{ij}$, $\mathbf{u}_{ij}$, and $T_{ij}$ are to impose that the
kinetic model reproduces the collisional equations of momentum and energy of the original Boltzmann operator
$J_{ij}^\el$, namely,
\begin{eqnarray}
\int\dd\mathbf{v}\, \mathbf{v} J_{ij}^\el[\mathbf{v}|f_i,f_j] &=&-\nu_{ij}\int \dd\mathbf{v}\,
\mathbf{v}\left[f_i(\mathbf{v})-f_{ij}(\mathbf{v})\right]\nn
&=&-\nu_{ij}n_i\left(\mathbf{u}_i-\mathbf{u}_{ij}\right), \label{4.13} \eeqa \beqa \int \dd\mathbf{v}\, v^2
J_{ij}^\el[\mathbf{v}|f_i,f_j] &=&-\nu_{ij}\int \dd\mathbf{v}\,
v^2\left[f_i(\mathbf{v})-f_{ij}(\mathbf{v})\right]\nn
&=&-\nu_{ij}n_i\left[\frac{d}{m_i}\left(\TT_i-T_{ij}\right)+u_i^2-u_{ij}^2\right]. \label{4.14b}
\end{eqnarray}
Conservation of total momentum in collisions $ij$ implies that
\begin{equation}
\nu_{ij}n_im_i\left(\mathbf{u}_i-\mathbf{u}_{ij}\right)+\nu_{ji}n_jm_j
\left(\mathbf{u}_j-\mathbf{u}_{ji}\right)=\mathbf{0}. \label{4.15}
\end{equation}
For physical reasons it is assumed that $n_i\nu_{ij}=n_j\nu_{ji}$, as happens with the effective collision
frequencies $\xi_{ij}$ defined by Eq.\ \eqref{5}. Moreover, the reference velocity $\mathbf{u}_{ij}$ is assumed
to be symmetrical, i.e., $\mathbf{u}_{ij}=\mathbf{u}_{ji}$. Under the above conditions, Eq.\ \eqref{4.15} yields
\begin{equation}
\mathbf{u}_{ij}=\mu_{ij}\mathbf{u}_{i}+\mu_{ji}\mathbf{u}_{j}. \label{4.16}
\end{equation}
To identify the remaining parameters $\nu_{ij}$ and $T_{ij}$ through
Eqs.\ \eqref{4.13} and \eqref{4.14b} we need to compute the
corresponding collision integrals associated with $J_{ij}^\el$. In
the original model proposed by Gross and Krook \cite{GK56}, the
left-hand sides of Eqs.\ \eqref{4.13} and \eqref{4.14b} were
evaluated by considering mixtures of Maxwell molecules, in which
case the collision rate is independent of the relative velocity.
Here, however, we want to keep in $J_{ij}^\el$ the velocity
dependence of the collision rate characteristic of hard spheres.

Applying the property \eqref{c1} and making use of Eqs.\ \eqref{c2} and \eqref{n6} with $\alpha_{ij}=1$, it is
easy to obtain
\begin{equation}
\int \dd\mathbf{v}_1 \,\mathbf{v}_1
J_{ij}^\el[\mathbf{v}_1|f_i,f_j]=-\frac{2\pi^{(d-1)/2}}{\Gamma((d+3)/2)}\mu_{ji} \sigma_{ij}^{d-1}n_in_j\langle
g_{12}\mathbf{g}_{12}\rangle_{ij}, \label{4.1}
\end{equation}
\begin{equation}
 \int \dd\mathbf{v}_1 \,{v}_1^2
J_{ij}^\el[\mathbf{v}_1|f_i,f_j]=-\frac{4\pi^{(d-1)/2}}{\Gamma((d+3)/2)}\mu_{ji} \sigma_{ij}^{d-1}n_in_j\langle
g_{12}\mathbf{g}_{12}\cdot \mathbf{G}_{12}\rangle_{ij},
\label{4.2}
\end{equation}
where the integrations over the solid angle have been performed. The averages appearing in Eqs.\ \eqref{4.1} and
\eqref{4.2} can be evaluated, as in the case of $\langle g_{12}^3\rangle_{ij}$, by assuming Gaussian forms for
$f_i$ and $f_j$ and neglecting terms of order third and higher in $\mathbf{u}_i-\mathbf{u}_j$. This is done in
Appendix \ref{appA} with the results \eqref{4.3} and \eqref{4.5}, so that Eqs.\ \eqref{4.1} and \eqref{4.2}
become
\begin{equation}
\int \dd\mathbf{v}_1 \,\mathbf{v}_1
J_{ij}^\el[\mathbf{v}_1|f_i,f_j]=-\xi_{ij}n_i\mu_{ji}(\mathbf{u}_{i}-\mathbf{u}_{j}), \label{4.4}
\end{equation}
\begin{eqnarray}
 \int \dd\mathbf{v}_1 \,{v}_1^2 J_{ij}^\el[\mathbf{v}_1|f_i,f_j]&=&-2\xi_{ij}n_i\mu_{ji}\left(\frac{\TT_i}
{m_i}+\frac{\TT_j}{m_j}\right)^{-1}
\left\{\left(\frac{\TT_j}{m_j}\mathbf{u}_i+\frac{\TT_i}{m_i}\mathbf{u}_j\right)
\cdot\left(\mathbf{u}_i-\mathbf{u}_j\right)\right.\nn &&\left. -\frac{d}{m_i+m_j}\left(\TT_j-\TT_i\right)
\left[\frac{\widetilde{T}_i}{m_i}+\frac{\widetilde{T}_j}{m_j}+\frac{3}{2d}
\left(\mathbf{u}_i-\mathbf{u}_j\right)^2\right]\right\},
\label{4.6}
\end{eqnarray}
 respectively, where
$\xi_{ij}$ is given by Eq.\ \eqref{5}. Substitution of Eqs.\ \eqref{4.4} and \eqref{4.6} into Eqs.\ \eqref{4.13}
and \eqref{4.14b} yields
\begin{equation}
 \nu_{ij}=\xi_{ij}, \label{4.17} \eeq \beq T_{ij}=\TT_i+\frac{2m_i
m_j}{(m_i+m_j)^2}\left\{\TT_j-\TT_i+\frac{(\mathbf{u}_i-\mathbf{u}_j)^2}{2d}\left[m_j+
\frac{\TT_j-\TT_i}{\TT_i/m_i+\TT_j/m_j}\right]\right\}. \label{4.18}
\end{equation}
Note that
$T_{ij}-\TT_i+T_{ji}-\TT_j=[m_im_j/(m_i+m_j)](\mathbf{u}_i-\mathbf{u}_j)^2/d$. It is also interesting to remark
that  the term proportional to $(\mathbf{u}_i-\mathbf{u}_j)^2(\TT_j-\TT_i)$ in Eq.\ \eqref{4.18} is absent in
the kinetic model based on the Boltzmann equation for Maxwell molecules \cite{GS03}.

Equations \eqref{4.16}, \eqref{4.17}, and \eqref{4.18}, along with
Eq.\ \eqref{4.12}, close the construction of the kinetic model
\eqref{4.10} for mixtures of elastic hard spheres. Next, the
corresponding kinetic model for granular mixtures is defined by Eq.\
\eqref{4.11}.
 This extended kinetic model has the same
 collisional transfer of momentum and energy as the true inelastic
 term $J_{ij}^\inel$, as a consequence of Eqs.\ \eqref{n2.7}, \eqref{n2.7bis}, \eqref{4.13}, and \eqref{4.14b},
 at least in the
 Gaussian approximation \eqref{A1}. As a consequence, the temperature ratios
 $\gamma_i=T_i/T$ of the HCS are the same as those obtained from the
 inelastic Boltzmann equation.

The NS transport coefficients predicted by the kinetic model can be
evaluated  from the Chapman--Enskog method. Since
$\mathbf{u}_i-\mathbf{u}=\mathbf{j}_i/m_in_i$, one has
\beq
\mathbf{u}_{ij}=\mathbf{u}+\frac{\mathbf{j}_i}{n_i(m_i+m_j)}+\frac{\mathbf{j}_j}{n_j(m_i+m_j)}.
\label{4.19}
\eeq
Furthermore, to NS order, $\TT_i=T_i=\gamma_i T$ and
\beq
T_{ij}=T\left[\gamma_i+\frac{2m_i
m_j}{(m_i+m_j)^2}\left(\gamma_j-\gamma_i\right)\right].
\label{4.20}
\eeq
 As a consequence, the reference distribution $f_{ij}$ becomes
\beq
f_{ij}(\mathbf{v})=f_{ij}^{(0)}(\mathbf{V})\left[1+\frac{\mu_{ij}}{T_{ij}}
\mathbf{V}\cdot\left(\frac{\mathbf{j}_i}{n_i}+\frac{\mathbf{j}_j}{n_j}
\right)\right],
\label{4.12bb}
\eeq
where
\beq
f_{ij}^{(0)}(\mathbf{V})=n_i\left(\frac{m_i}{2\pi
T_{ij}}\right)^{d/2} \exp\left(-\frac{m_i}{2T_{ij}}V^2\right).
\label{4.12bbb}
\eeq
Therefore, in the binary case, the linear operators $\mathcal{L}_1$
and $\mathcal{M}_1$ take the forms
\beqa
{\cal
L}_{1}f_{1}^{(1)}&=&\frac{1+\alpha_{11}}{2}\nu_{11}\left(f_{1}^{(1)}-f_{11}^{(0)}
\mathbf{V}\cdot\frac{\mathbf{j}_1}{n_1\gamma_1
T}\right)+
\frac{1+\alpha_{12}}{2}\nu_{12}\left(f_{1}^{(1)}-f_{12}^{(0)}\frac{\mu_{12}}{T_{12}}
\mathbf{V}\cdot\frac{\mathbf{j}_1}{n_1}\right)\nn
&& -\frac{\zeta_{11}+\zeta_{12}}{2}\left(\frac{\partial}{\partial
{\bf v}}\cdot {\bf V}f_1^{(1)}- \frac{{\bf
j}_1}{\rho_1}\cdot\frac{\partial}{\partial {\bf v}}f_1^{(0)}\right),
\label{4.21}
\eeqa
\begin{equation}
{\cal
M}_{1}f_{2}^{(1)}=-\frac{1+\alpha_{12}}{2}\nu_{12}f_{12}^{(0)}\frac{\mu_{12}}{T_{12}}
\mathbf{V}\cdot\frac{\mathbf{j}_2}{n_2}.
\label{4.22}
\end{equation}
The transport coefficients are given again by Eqs.\
\eqref{a13}--\eqref{a15},   \eqref{3.17}, \eqref{3.18}, and
\eqref{4.7}--\eqref{4.9}, but now the associated collision
frequencies \eqref{a10}, \eqref{b13}, \eqref{b14}, and
\eqref{4.17b}--\eqref{4.19b} are computed by using the linear
operators \eqref{4.21} and \eqref{4.22}. In particular, the
transport coefficients associated with the mass flux are the same as
those obtained from the inelastic Boltzmann equation
\cite{GD02,GM06}.

\end{document}